\documentclass[fleqn,usenatbib]{mnras}

\usepackage{newtxtext,newtxmath}

\usepackage[T1]{fontenc}
\usepackage{ae,aecompl}
\usepackage{ragged2e}

\DeclareRobustCommand{\VAN}[3]{#2}
\let\VANthebibliography\thebibliography
\def\thebibliography{\DeclareRobustCommand{\VAN}[3]{##3}\VANthebibliography}

\usepackage{graphicx}	
\usepackage{amsmath}	

\usepackage{multicol}   
\usepackage{bm}		    
\usepackage{pdflscape}	
\usepackage{xcolor}	    

\usepackage{gensymb}

\title[Spitzer Legacy for Debris Disk Variability]{Activity in White Dwarf Debris Disks I:  {\em Spitzer} Legacy Reveals Variability Incompatible with the Canonical Model}

\author[H.~T.~Noor et al.]{Hiba Tu Noor,$^{1}$\thanks{E-mail: hiba.noor.19@ucl.ac.uk} 
Jay Farihi,$^{1}$
Scott J.~Kenyon,$^{2}$
Roman R.~Rafikov,$^{3}$
Mark C.~Wyatt,$^{4}$
Kate Y.~L.~Su,$^{5}$
\newauthor
Carl Melis,$^{6}$
Andrew Swan,$^{7}$
Thomas G.~Wilson,$^{7}$
Boris T.~G\"{a}nsicke,$^{7}$
Amy Bonsor,$^{4}$
Laura K.~Rogers,$^{8}$
\newauthor
Seth Redfield,$^{9}$
and Mukremin Kilic$^{10}$
\\
$^{1}$Department of Physics and Astronomy, University College London, London WC1E 6BT, UK\\
$^{2}$Smithsonian Astrophysical Observatory, Cambridge MA 02138, USA\\
$^{3}$Department of Applied Mathematics and Theoretical Physics, University of Cambridge, Cambridge CB3 0WA, UK\\
$^{4}$Institute of Astronomy, University of Cambridge, Madingley Road, Cambridge CB3 0HA, UK\\
$^{5}$Space Science Institute, Boulder CO 80301, USA\\
$^{6}$Astronomy \& Astrophysics Department, University of California, San Diego CA 92093-0424, USA\\
$^{7}$Department of Physics, University of Warwick, Coventry CV4 7AL, UK\\
$^{8}$NSF National Optical-Infrared Astronomy Research Laboratory, 950 North Cherry Avenue, Tucson, AZ 85719, USA\\
$^{9}$Astronomy Department and Van Vleck Observatory, Wesleyan University, Middletown, CT 06459, USA\\
$^{10}$Homer L. Dodge Department of Physics and Astronomy, University of Oklahoma, 440 W. Brooks St., Norman, OK 73019, USA\\
}

\date{Accepted 2025 August 4. Received 2025 July 29; in original form 2025 May 6}

\pubyear{2025}

\begin{document}
\label{firstpage}
\pagerange{\pageref{firstpage}--\pageref{lastpage}}
\maketitle

\begin{abstract}
This study presents all available, multi-epoch 3.6 and 4.5\,$\upmu$m photometry from {\it Spitzer Space Telescope} observations of white dwarf debris disks, including weekly cadence observations of 16 relatively bright systems, and 5\,h staring-mode observations for five of these.  Significant variability is detected in 85\,per cent of disks and across all timescales probed, from minutes to weeks to years, where the largest flux changes correlate with the longest time baselines, and the infrared excesses persist utterly.  While each source is idiosyncratic, the overall results indicate the most variable disks correlate with those that are the brightest (dustiest), and also among those with detected gas, demonstrating both dust and gas are produced via ongoing collisions. There is a correlation between flux and colour changes, where disks tend to appear redder when dimmer and bluer when brighter, consistent with an excess of small dust grains produced in collisions, followed by a gradual return to equilibrium. The overall results are a drastic departure from the predictions of the canonical {-- geometrically thin, optically thick --} disk in both flux and colour, but are broadly consistent with collisional evolution based on a simple model. The data presented herein constitute a legacy resource that can inform time-series studies of polluted and dusty white dwarfs, and importantly serve as a basis for future disk modelling, beyond the pioneering canonical framework.

\end{abstract}

\begin{keywords}
circumstellar matter -- planetary systems -- white dwarfs
\end{keywords}



\section{Introduction}

Planetesimal disks are a ubiquitous feature seen during various stages of stellar evolution, from their initial emergence around young main-sequence stars to their persistence or regeneration around white dwarfs in the post-main sequence. While dusty disks around young stars are critical for understanding planet formation and migration, debris disks around mature main-sequence stars are second-generation structures, produced through collisional grinding of large planetesimals (see \citealt{Wyatt_2008, Najita_2022}, and references therein). Many of these debris disks contain $T\lesssim100$\,K material that orbits at tens to hundreds of astronomical units \citep[e.g.][]{Hughes_2018}, although some also exhibit mid-infrared excesses, indicating the presence of warmer dust nearer to the star \citep{Ballering_2014}.  This warmer dust can form within an inner planetesimal belt, or may originate from and be sustained by material transported from an outer reservoir \citep[e.g.][]{KennedyPiette_2015, Marboeuf_2016}.  Debris disks around white dwarfs are modeled as the result of the latter process, where planetesimals are dynamically transported inwards, and subsequently tidally disrupted \citep{Jura_2003} at the stellar Roche limit, leading to the formation of a relatively compact debris disk {with a typical radius of order 1\,$R_{\odot}$} \citep[for a review, see][]{Farihi_2016}.

Approximately 2\,per cent of white dwarfs show detectable infrared excesses consistent with the presence of $T\simeq 1000$\,K, dusty debris disks \citep{Wilson_2019}, which is an order of magnitude smaller than the fraction of white dwarfs harbouring planetary systems as indicated by photospheric metal pollution (over 25\,per cent; \citealt{Zuckerman_2010, Koester_2014}).  Roughly 60 debris disks have now been detected orbiting white dwarfs{, all of which also exhibit atmospheric metal pollution,} with over 100 candidate detections \citep[e.g.][]{Dennihy_2020a, Lai_2021}. A subset of dust disks also exhibit a gaseous component that is detected through emission lines originating in the same region as the disk solids \citep[e.g.][]{Gansicke_2006, Melis_2010, Melis_2020, Dennihy_2020b, GentileFusillo_2021b}, or via absorption lines along the line of sight \citep[e.g.][]{Gansicke_2012, Xu_2016}.

Observations have been complemented by theoretical efforts to model debris disks around white dwarfs \citep[e.g.][]{Rafikov_2011, Metzger_2012, KenyonBromley_2017a, Lieshout_2018, Malamud_2020a}.  For over a decade, the standard model was a circular, geometrically thin, optically thick disk analogous to the rings of Saturn, and located entirely within the Roche limit (hereafter, the canonical model, \citealt{Jura_2003, Rafikov_2011}).  However, there is now significant evidence to indicate that this model is insufficient to account for a range of observations. For example, many infrared variations cannot be reconciled with a flat disk of fixed radial extent \citep{Swan_2020}, the infrared brightnesses of some disks are too large to be reproduced by the canonical model \citep{Jura_2007,Dennihy_2017,Farihi_2018, GentileFusillo_2021b}, and in the case of SDSS\,J1557+0916, the model is dynamically prohibited \citep{Farihi_2017}. Thus, while the canonical model provided an important foundational framework, it is increasingly clear that alternative configurations must be considered, underscoring the need for new observational data.

In debris disks around main-sequence stars, infrared variability is often leveraged to monitor planetary dynamics and can identify significant collisional events manifesting as extreme infrared excesses \citep[e.g.][]{Melis_2012, Meng_2012, Su_2019, Rieke_2021}. White dwarf debris disks exhibit similar phenomena, with several published examples where unambiguous 3--5\,$\upmu$m variability has been observed over months to years \citep[][]{Xu_2018a, Farihi_2018, Wang_2019, Swan_2019, Swan_2020, Guidry_2024}.  In contrast, little is known about variability on shorter timescales, or at wavelengths longer than those accessible to relatively recent, post-cryogenic, infrared space missions.  

The {\it Wide-field Infrared Survey Explorer (WISE)} satellite, which provided mid-infrared data biannually until late 2024, remains a valuable resource for investigating infrared variability on timescales ranging from months to years \citep{Swan_2019}. Notably, however, the orbital periods, and thus collision timescales, for bodies near to or within the Roche radius are on the order of hours, and thus there is a three orders-of-magnitude gap between the relevant disk orbital periods and the sampling cadence of {\em WISE}. Fortunately, legacy {\em Spitzer Space Telescope} observations are available to close this gap in observing cadence, and together with {\it WISE} data, both short- and long-term variations can be characterised.

This study is part of an investigation aimed at constraining the timescales and amplitudes of changes in dusty white dwarf disk emission. The current paper utilises 3.6 and 4.5\,$\upmu$m data from the cryogenic and warm {\it Spitzer} mission to analyse variability on timescales ranging from minutes to weeks.  A forthcoming work will extend this investigation to longer timescales based on the completed {\it WISE} mission (Noor et al.\ in preparation). Section~\ref{sec:Observations} describes the observations and data reduction.  Section~\ref{sec:Results} presents the results of the survey, discusses flux changes over a range of timescales, and characterises the observed variability at the population level. Section~\ref{sec:Discussion} explores the likely causes of the observed variability, and applies a simple collisional cascade model to each system with a sufficient number of observations. A summary is provided in Section~\ref{sec:Conclusions}.

\begin{table*}
\begin{center}	 
\caption{White dwarfs with infrared excesses consistent with dusty debris disks in this study.}
\label{tab:targets}
\begin{tabular}{l@{\enspace}l@{}c@{\quad}ccr@{\enspace}lr@{\enspace}lrccl} 
\hline

WD      
&Alternate name   &Ca\,{\sc ii}       
&\multicolumn{2}{c}{$N$ epochs}   
&\multicolumn{4}{c}{$|\Delta F_{\rm max}|$ (\%)} 
&${T_{\rm dust}}$         &${R_{\rm dust}}$   
&$L_{\rm dust}/L_\star$                 
&Ref.\\

&                         
&emission           
&3.6\,$\upmu$m  
&4.5\,$\upmu$m           
&\multicolumn{2}{@{\hskip -4pt}c}{3.6\,$\upmu$m}   
&\multicolumn{2}{@{\hskip -4pt}c}{4.5\,$\upmu$m}  
&(K)&($R_{\odot}$)    
&(\%)  
&\\
       
\hline

\multicolumn{3}{l}{Targets in programme 14258:}         &    &     &       &                   &       &                    &       &       &      &\\
0110$-$565            &                            &    &19  &19   &27.5   &($24\,\upsigma$)   &29.8   &($25\,\upsigma$)    &1160   &1.68   &0.22  &1\\
0146+187              &                            &    &14  &14   &6.6   &($6.8\,\upsigma$)  &5.0   &($6.4\upsigma$)     &980    &0.92   &1.43  &2\\
0300$-$013            &GD\,40                      &    &14  &14   &19.9   &($27\,\upsigma$)   &22.1   &($31\,\upsigma$)    &1130   &0.91   &0.42  &3\\
0408$-$041  &GD\,56                      &    &20  &21   &31.7   &($67\,\upsigma$)   &28.3   &($52\,\upsigma$)    &980    &1.59   &2.31  &4\\
0420+520              &                            &    &10  &9    &20.8   &($14\,\upsigma$)   &16.0   &($9.1\,\upsigma$)   &1340   &1.68   &0.23  &5\\
0420$-$731            &                            &    &29  &29   &9.7   &($7.4\,\upsigma$)  &10.8   &($8.7\,\upsigma$)   &1140   &1.67   &0.58  &5 \\
0435+410              &GD\,61                      &    &11  &12   &11.8   &($9.3\,\upsigma$)  &10.6 &($7.7\,\upsigma$)   &1520   &0.72   &0.28  &6\\
0842+231              &Ton\,345                    &+   &13  &14   &17.6   &($15\,\upsigma$)   &12.4   &($12\,\upsigma$)    &1440   &1.13   &0.66  &7, 8\\
0842+572    &                            &+   &13  &16   &13.6   &($15\,\upsigma$)   &11.3   &($11\,\upsigma$)    &1210   &1.18   &7.17  &9\\
0843+516              &                            &    &14  &14   &7.3   &($5.5\,\upsigma$)  &29.4   &($23\,\upsigma$)    &1370   &1.98   &0.21  &6\\
1226+110    &SDSS\,J122859.93+104032.9   &+   &15  &16   &37.5   &($27\,\upsigma$)   &39.7   &($24\,\upsigma$)    &1150   &1.92   &0.44  &10, 11\\
1536+520              &                            &    &28  &28   &19.0   &($28\,\upsigma$)   &9.8   &($19\,\upsigma$)    &1110   &2.02   &1.85  &12\\
1729+371              &GD\,362                     &    &34  &35   &26.7   &($30\,\upsigma$)   &11.2   &($17\,\upsigma$)    &960    &0.70   &2.28  &13\\
2115$-$560  &                            &    &18  &22   &7.7   &($14\,\upsigma$)   &9.3   &($23\,\upsigma$)    &910    &0.72   &0.76  &14\\
2221$-$165            &                            &    &14  &14   &28.1   &($16\,\upsigma$)   &27.6   &($13\,\upsigma$)    &1120   &0.46   &0.81  &15\\
2329+407   &                            &    &13  &14   &16.9   &($35\,\upsigma$)   &26.6   &($56\,\upsigma$)    &790    &2.94   &0.26  &5\\
\hline
\multicolumn{3}{l}{Other targets in this study:}        &     &    &      &                    &       &                    &       &       &      &\\
J0006+2858            &SDSS\,J000634.72+285846.5   &+   &2    &3   &4.8  &($1.4\,\upsigma$)   &14.6   &($5.0\,\upsigma$)   &990    &3.68   &0.28  &9, 16\\
0106$-$328            &                            &    &2    &3   &30.8  &($3.4\,\upsigma$)   &33.3   &($4.8\,\upsigma$)   &1320   &0.94   &0.09  &2\\
0145+234              &                            &+   &14   &15  &20.2  &($33\,\upsigma$)    &22.2  &($22\,\upsigma$)    &1260   &0.65   &1.23  &9, 17\\
J0234$-$0406          &SDSS\,J023415.51$-$040609.1 &+   &2    &2   &21.3  &($4.5\,\upsigma$)     &10.4   &($2.7\,\upsigma$)     &1200   &0.96   &0.60  &16,18\\
0246+734              &                            &    &2    &3   &13.2  &($1.4\,\upsigma$)   &39.1   &($6.0\,\upsigma$)   &1300   &0.20   &0.73  &19\\
0307+077              &                            &    &2    &2   &24.4  &($0.8\,\upsigma$)   &59.0   &($3.7\,\upsigma$)   &1070   &0.57   &0.11  &15\\
J0738+1835            &SDSS\,J073842.56+183509.6   &+   &3    &3   &32.6  &($11\,\upsigma$)    &33.1   &($12\,\upsigma$)    &1300   &0.73   &2.20  &20\\
J0959$-$0200          &SDSS\,J095904.69$-$020047.6 &+   &3    &3   &42.2  &($15\,\upsigma$)    &37.6   &($14\,\upsigma$)    &1180   &0.76   &2.60  &21, 22\\
1015+161              &                            &    &2    &3   &10.1  &($2.3\,\upsigma$)   &12.0   &($3.1\,\upsigma$)   &1330   &1.33   &0.22  &3\\
1018+410              &                            &    &3    &3   &14.1  &($3.1\,\upsigma$)   &12.9   &($3.4\,\upsigma$)   &1260   &1.82   &0.13  &23\\
1041+091              &SDSS\,J104341.53+085558.2   &+   &2    &2   &12.2  &($2.9\,\upsigma$)   &19.7   &($5.0\,\upsigma$)   &1500   &0.81   &0.26  &24, 25\\
1116+026              &GD\,133                     &    &2    &2   &3.1  &($0.5\,\upsigma$)  &2.2   &($0.5\,\upsigma$)  &1140   &0.75   &0.55  &3\\
1145+017              &                            &    &1    &10  &      &                    &22.0   &($6.4\,\upsigma$)   &1320   &0.79   &0.40  &26, 27\\
1150$-$153            &                            &    &5    &5   &6.4  &($3.1\,\upsigma$)   &8.7   &($3.9\,\upsigma$)   &910    &1.05   &2.05  &28\\
J1221+1245$^{\rm c}$  &SDSS\,J122150.81+124513.3   &    &2    &2   &1.3  &($0.5\,\upsigma$)     &0.9   &($0.4\,\upsigma$)     &1210   &0.60   &3.49  &22\\
1232+563              &                            &    &2    &2   &2.3  &($0.7\,\upsigma$)  &9.4   &($3.1\,\upsigma$)   &1020   &0.90   &2.04  &12\\
1349$-$230            &                            &+   &1    &2   &      &                    &16.8   &($5.2\,\upsigma$)   &1260   &1.17   &0.33  &1, 29\\
1457$-$086            &                            &    &2    &2   &3.5  &($0.4\,\upsigma$)  &4.3   &($0.6\,\upsigma$)  &1350   &1.48   &0.05  &2\\
1541+650              &                            &    &2    &2   &10.1  &($2.6\,\upsigma$)   &12.7   &($3.9\,\upsigma$)   &860    &1.11   &1.20  &30\\
1551+175              &                            &    &2    &2   &11.7  &($1.1\,\upsigma$)   &19.3   &($2.3\,\upsigma$)   &820    &2.25   &0.09  &19\\
1554+094              &                            &    &4    &4   &37.9  &($13\,\upsigma$)    &35.5   &($13\,\upsigma$)    &920    &5.30   &0.34  &22\\
J1617+1620            &SDSS\,J161717.04+162022.4   &+   &2    &2   &13.8  &($3.9\,\upsigma$)   &7.1   &($2.0\,\upsigma$)   &1630   &0.36   &1.92  &7, 31\\
J1931+0117            &GALEX\,J193156.8+011745     &    &2    &2   &6.6  &($1.8\,\upsigma$)   &5.1   &($1.5\,\upsigma$)   &1290   &2.00   &0.22  &32\\
2132+096              &                            &    &2    &2   &19.1  &($1.5\,\upsigma$)   &4.5   &($0.6\,\upsigma$)  &920    &1.35   &0.11  &19\\
2207+121              &                            &    &2    &2   &9.4  &($2.7\,\upsigma$)   &3.9   &($1.2\,\upsigma$)   &1130   &1.09   &0.56  &33\\
2326+049              &G29-38                      &    &7    &6   &15.3  &($29\,\upsigma$)    &9.1   &($19\,\upsigma$)   &1010   &0.78   &2.60  &34\\
\hline
\end{tabular}
\end{center}
\justifying
\noindent
{\em References:}
(1) \citet{Girven_2012}; (2) \citet{Farihi_2009}; (3) \citet{Jura_2007}; (4) \citet{Kilic_2006}; (5) \citet{Hoard_2013}; (6) \citet{Farihi_2011}; (7) \citet{Brinkworth_2012}; (8) \citet{Gansicke_2008}; (9) \citet{Melis_2020}; (10) \citet{Gansicke_2006}; (11) \citet{Brinkworth_2009}; (12) \citet{Debes_2011}; (13) \citet{Becklin_2005}; (14) \citet{Mullally_2007}; (15) \citet{Farihi_2010}; (16) \citet{Lai_2021}; (17) \citet{RebassaMansergas_2019}; (18) \citet{GentileFusillo_2021b}; (19) \citet{Bergfors_2014}; (20) \citet{Dufour_2010}; (21) \citet{Girven_2011}; (22) \citet{Farihi_2012}; (23) \citet{Rocchetto_2015}; (24) \citet{Gansicke_2007}; (25) \citet{Melis_2010}; (26) \citet{Vanderburg_2015}; (27) \citet{Cauley_2018}; (28) \citet{KilicRedfield_2007}; (29) \citet{Melis_2012}; (30) \citet{Kilic_2012}; (31) \citet{Wilson_2014}; (32) \citet{Melis_2011}; (33) \citet{XuJura_2012}; (34) \citet{ZuckermanBecklin_1987}.
\end{table*}

\section{Observations and Data Reduction}
\label{sec:Observations}

\subsection{Dusty disk sample}

The {\it Spitzer} Heritage Archive was searched for archival, Infrared Array Camera \citep[IRAC;][]{Fazio_2004} observations of all white dwarfs with known circumstellar debris disks. Within the data available, the largest subset were observed as part of programme 14258 (PI Farihi), which created a sizable legacy for dusty white dwarf variability, with approximately 87\,h of dedicated, multi-epoch observations of 16 relatively bright targets known at that time, carried out between 2019 Jun and 2020 Jan. This programme included 12 logarithmically-sampled visits over 40\,d for each of the 16 stars, and between three to 16 linearly-sampled visits over 200\,d for four sources chosen for their long visibility windows. Measurements were obtained using the medium step size, cycling dither pattern in both channels 1 and 2 at 3.6 and 4.5\,$\upmu$m. For four of the 16 white dwarfs, staring-mode observations were also performed at 4.5\,$\upmu$m. Exposure times for all observations in this programme ranged from 12 to 30\,s, depending on target brightness, with the goal of achieving a signal-to-noise ratio (S/N) of at least 50 per visit in both channels. Archival data from the cryogenic mission in channels 3 and 4, at 5.7 and 7.9\,$\upmu$m, are not analysed owing to sparse coverage.

Among the 62 known dusty disks, 42 systems possess at least two epochs of IRAC observations in at least one channel, and these form the subset used for subsequent analysis.  The data analysed here thus include those examined in \citet{Swan_2020}, with the addition of higher-cadence observations from programme 14258, the staring-mode observations of the white dwarf 1226+110 (programme 14274; PI Wilson), and bi-daily observations of 0145+234 during the end of its outburst (programme 14322; PI Swan), contributing an additional 92\,h of monitoring beyond what has been previously published.

Table~\ref{tab:targets} lists the 42 dusty white dwarfs in the sample, together with the number of observations in each of the IRAC bandpasses. A full observation log is provided in Appendix~\ref{sec:obs_log}, including dates of individual pointings in each bandpasses, and the total fluxes measured in this work for all 62 known dusty white dwarfs.

\begin{figure*}
\includegraphics[width=\textwidth]{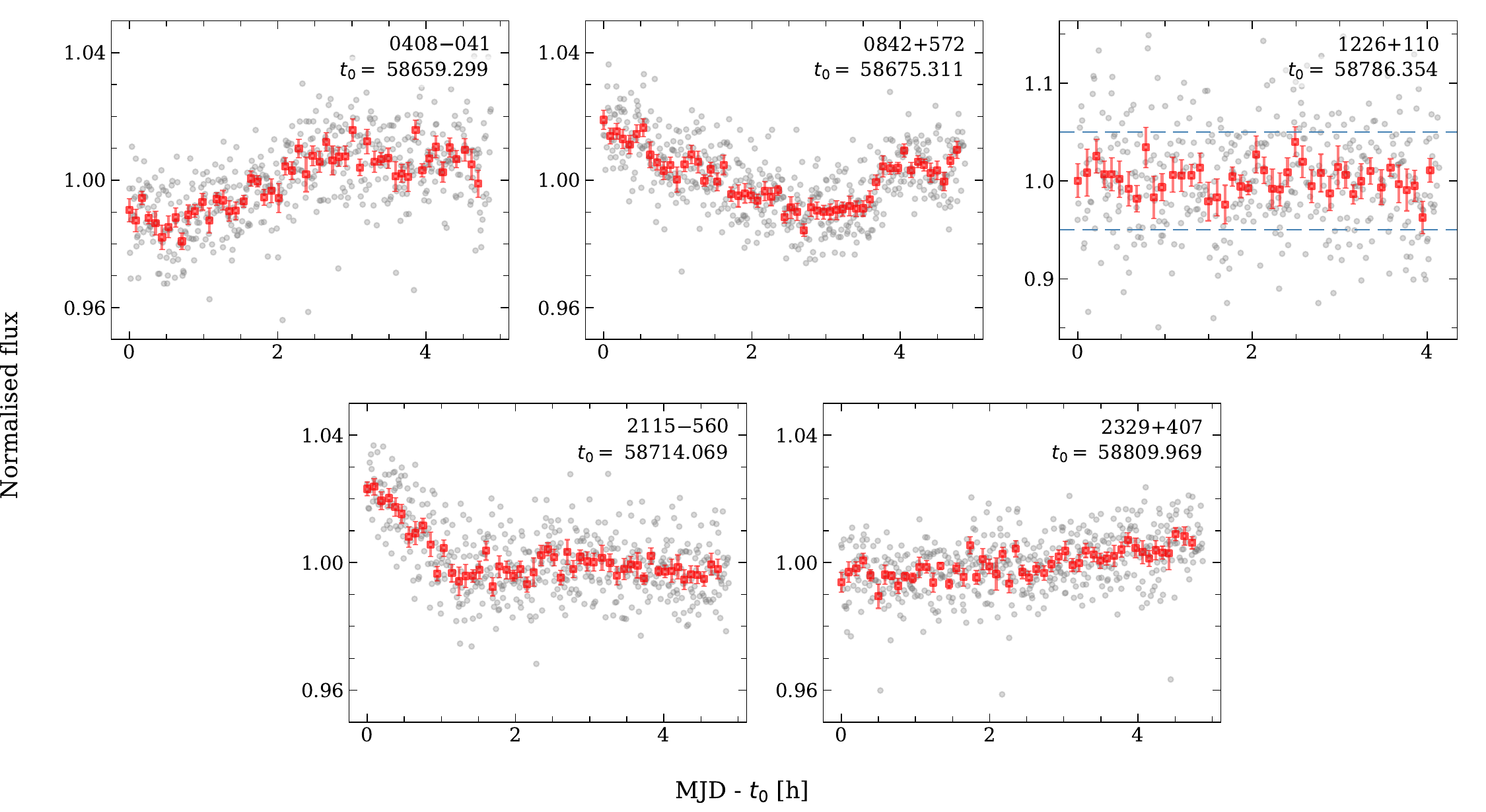}
\vskip -3pt
\caption{Normalised photometry of the five target stars obtained using IRAC staring-mode observations at 4.5\,$\upmu$m. Individual measurements are depicted by grey circles where each corresponds to a single 30\,s exposure, while red squares represent data that have been binned by a factor of 10 (i.e.\ the weighted average over 5\,min). Uncertainties in the binned fluxes were calculated using the standard error of the mean and do not include calibration error. The x-axis shows the time in hours offset by a reference time $t_0$, at the start of each observation. The dashed lines on the plot for 1226+110 represent the y-axis range used for the other four targets. The disks orbiting 0842+572 and 1226+110 show Ca \textsc{ii} line emission, while the other three do not.}

\label{fig:stare1}
\end{figure*}

\subsection{Standard imaging photometry}

The observational data can be divided into two types, standard observations using dithering patterns, and those executed in staring mode.  For standard observations, individual basic calibrated data frames were downloaded from the archive and combined to create science-grade mosaics with 0.6\,arcsec\,pixel$^{-1}$ using the \textsc{mopex} package, following published observatory guidelines and recommendations\footnote{\url{https://irsa.ipac.caltech.edu/data/SPITZER/docs/spitzermission/}}.  Aperture photometry was conducted on both the \textsc{mopex} mosaics using the \textsc{apex} module, and on the pipeline mosaics using the \textsc{apphot} package within \textsc{iraf}, using aperture radii of 2--5 native pixels (2.4--6.0\,arcsec) depending on target brightness, and sky annuli with 12--20 native pixels (14.4--24.0\,arcsec). 

Aperture and array-location-dependent corrections were applied, but no colour corrections were implemented. The bulk of targets in both bandpasses have photometric S/N $>100$, and thus negligible contribution towards the uncertainties in absolute flux.  The total errors in absolute brightness are thus limited by the instrument calibration uncertainty, which is $\simeq 2$\,per cent in all four IRAC channels \citep{Reach_2005}.  However, most of the subsequent analysis relies on relative flux changes, and for those, the calibration uncertainty is irrelevant.  The different methods of measuring flux -- aperture photometry on the \textsc{mopex} and pipeline mosaics -- were typically found to agree within 1\,per cent. The largest differences between the two methods can be up to 10\,per cent, but are typically owing to inadequate masking of cosmic rays in the pipeline mosaics. 

Note that \textsc{apex} performs both aperture photometry and point spread function (PSF) fitting. While the former is preferred owing to the undersampled PSF in IRAC, the latter can be useful in crowded fields if contamination from neighbouring sources is suspected. PSF fitting was thus explored for targets where such contamination was evident. However, \textsc{apex} often failed to identify the relevant background sources. In these cases, therefore, aperture photometry was retained using \textsc{iraf} \textsc{apphot}, with careful centroid placement and a fixed aperture radius of 2 native pixels to minimise flux contamination from nearby objects.

To best constrain the significance of flux changes at all targets, differential photometry was conducted.  For this purpose, well-isolated, non-varying stars with fluxes similar to or greater than that of the science target were selected (only those in the linear accumulation regime), and aperture photometry was performed as previously described. {At each epoch, the weighted mean flux of all selected comparison stars was calculated, and the flux of the science target at that epoch was divided by this mean value to obtain a relative flux. These relative fluxes were then normalised and used to quantify disk flux variability}. The errors in the individual relative fluxes were calculated by adding the measurement errors of the comparison and target stars in quadrature.

\subsection{Observations in IRAC staring mode}

Observations for five dusty white dwarfs were planned following the observatory recommendations for designing high-precision, time-series photometry.  Each pointing consisted of approximately 5\,h of continuous monitoring on-source, covering roughly one orbital period at the Roche radius. These staring mode observations were successfully carried out for the targets $0408-041$, $0842+572$, $1226+110$, $2115-560$, and $2329+407$.  None of these stars are known to be obscured by transiting debris \citep{Robert_2024}, ensuring that any observed variability cannot be caused by occultations. The pixel-phase effect, primarily caused by intra-pixel quantum efficiency variations, can result in flux variations based on the location of a point source within a pixel, and represents the largest source of correlated noise in IRAC photometry \citep{Ingalls_2012}.  To minimise this correlated noise, a point source can be placed on an area of the detector with minimal gain variation, otherwise known as the `sweet spot'. All staring-mode observations were thus designed to keep the target within this region.

Prior and subsequent to the science target observations, a 25\,min series of 30\,s calibration frames were taken to account for the spacecraft initial pointing drift \citep{Grillmair_2012}, as well as to obtain an estimate of the dark and sky backgrounds, using the same setup as the science frames. The corrected basic calibrated data (CBCD) files were then median combined to create a localised dark frame, which represents a measurement of the dark current and an estimate of the sky background in the region where the science target is located. This frame was subsequently subtracted from each of the science CBCD frames. Aperture photometry was then conducted on the resulting frames as previously described, excluding a few frames where cosmic rays were present in the region of the target. In all observations, the science target remained within 0.2 pixels of the sweet spot, achieving a typical S/N $\approx150$ per frame. As the last step, the \textsc{idl} programme \textsc{iracpc\_pmap\_corr}  was used to correct for intra-pixel variability.

\section{Results}
\label{sec:Results}

For both standard imaging photometry and staring-mode observations, the flux contribution from the dust was isolated from the total measured flux as follows.  Optical photometry for each target was obtained from catalogues such as Pan-STARRS \citep{Chambers_2016} and fitted with a pure hydrogen or helium white dwarf atmosphere model of appropriate effective temperature based on the literature \citep{Koester_2010}.  The stellar contribution, extrapolated to the appropriate wavelength, and calculated using synthetic photometry within each IRAC filter, was then subtracted from the total observed flux.

The light curves resulting from the multi-epoch, standard IRAC imaging of the 16 targets with short-term monitoring in programme 14258 are plotted in Appendix~\ref{sec:lightcurves}.  These plots include all available data sets during both the cryogenic and warm missions, together with a zoomed panel for additional detail on the numerous, most recent flux measurements.  The resulting light curves for the five staring-mode observations are shown in Figure~\ref{fig:stare1}.  For all light curves, normalisation was achieved by dividing each flux ratio by the mean.  The multi-epoch flux densities for each star in each IRAC channel are given in Appendix~\ref{sec:obs_log}: these are the values determined by aperture photometry, prior to separating the contributions of the disk and the stellar photosphere, as described above.  In this work, flux changes are calculated as percentages relative to the earlier epoch, i.e.\ $\Delta F = [(F_1 - F_2)/F_1] \times 100$.

\begin{figure}
\includegraphics[width=\columnwidth]{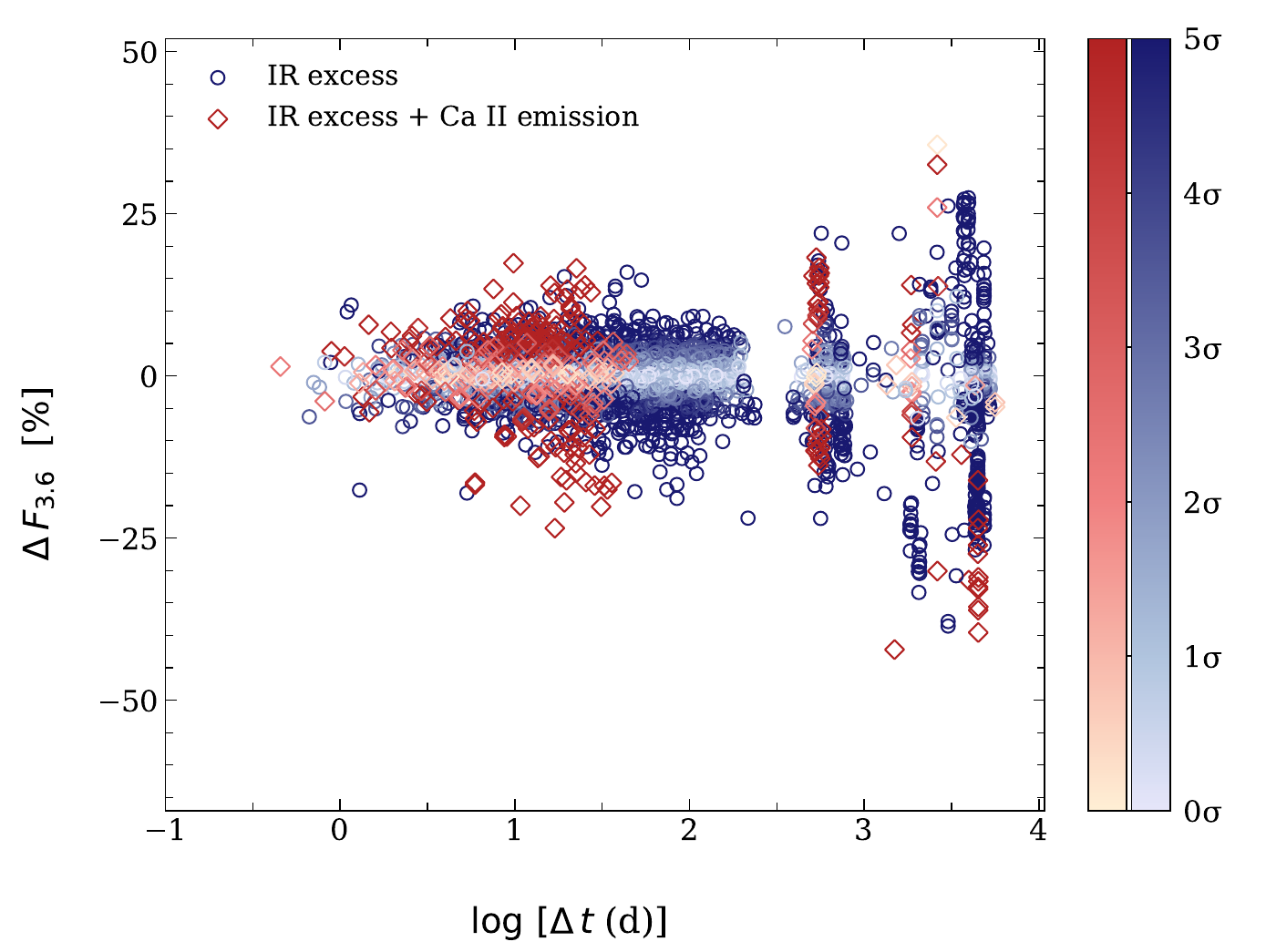}
\includegraphics[width=\columnwidth]{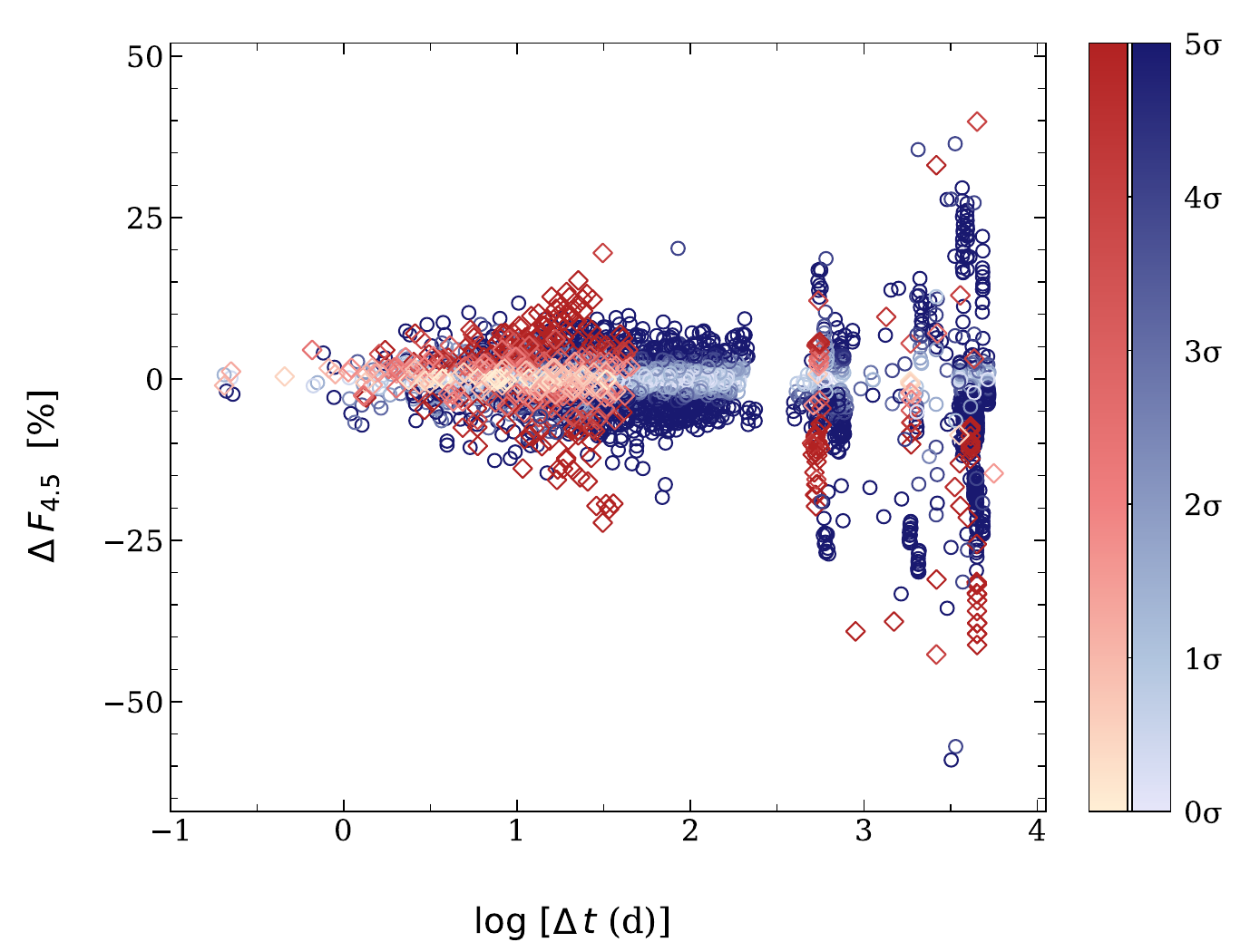}
\includegraphics[width=\columnwidth]{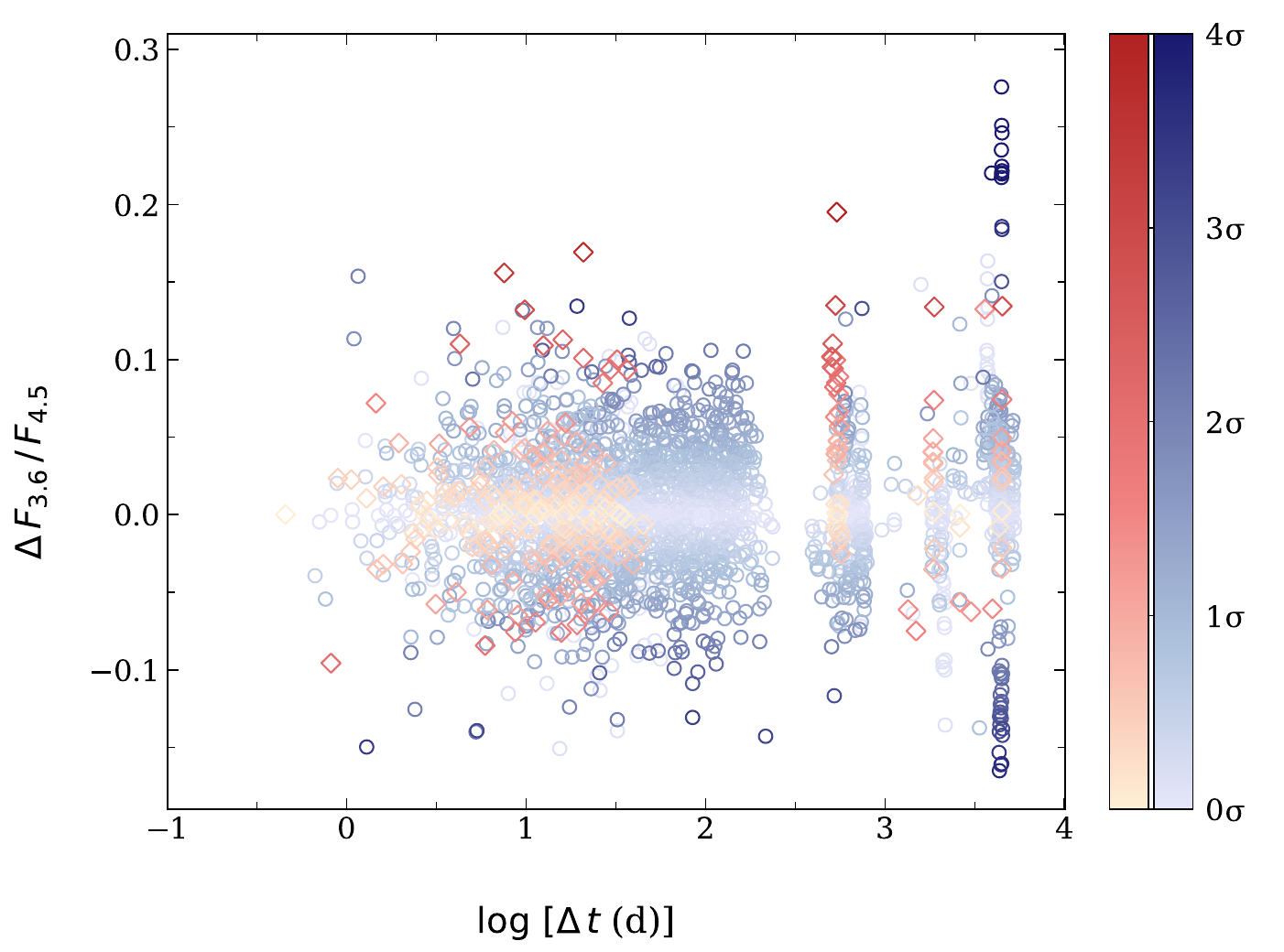}
\vskip -3pt
\caption{Flux changes at 3.6 and 4.5\,$\upmu$m, as well as colour changes for each pair of measurement across all targets with at least two epochs of observations. The colour bar indicates the significance of the changes, with darker symbols corresponding to higher levels of significance.  Debris disks with detected gas emission are indicated by red diamonds, while those with only infrared emission are plotted as blue circles.}
\label{fig:trends1}
\end{figure}

\begin{figure}
\includegraphics[width=\columnwidth]{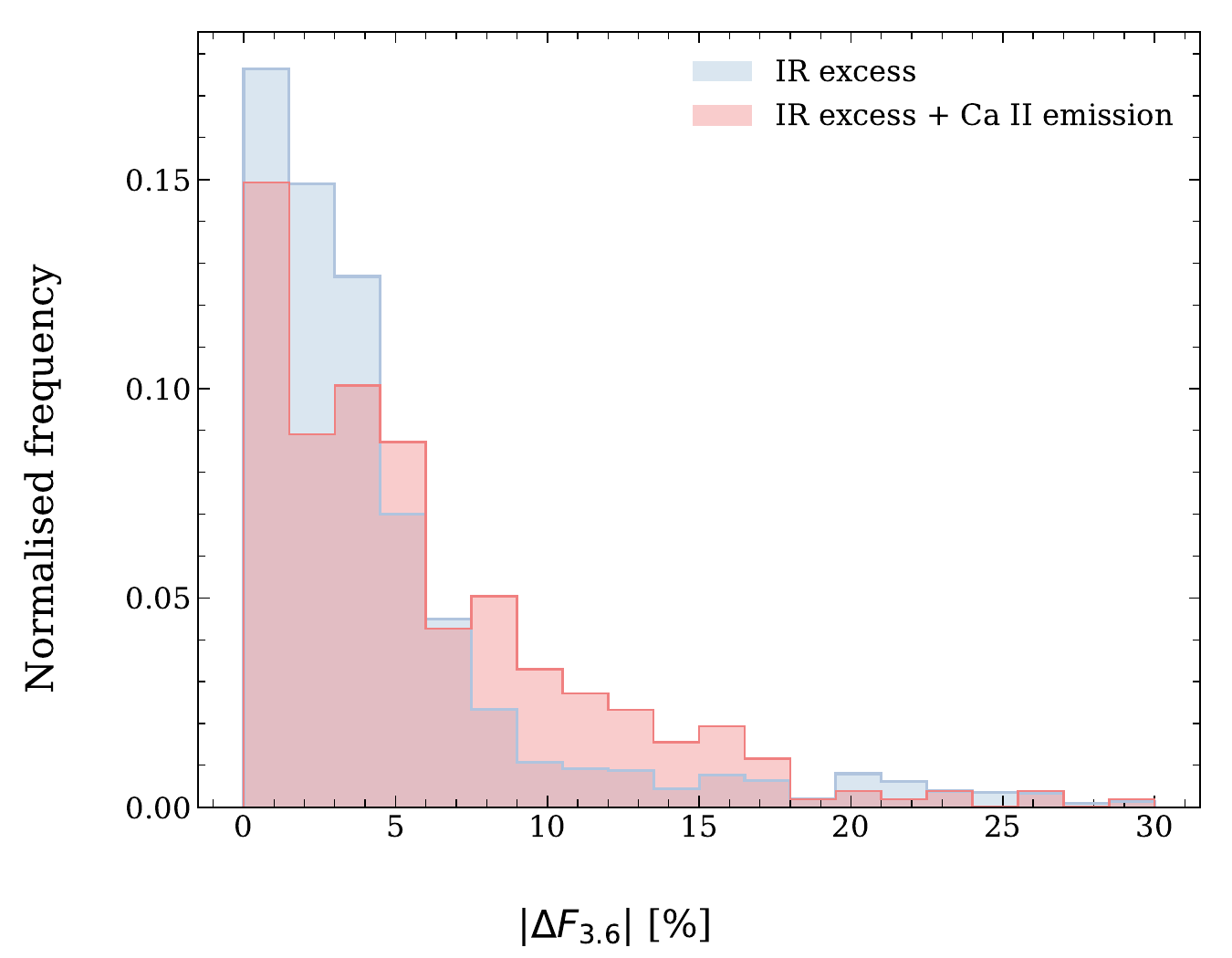}
\vskip -3pt
\caption{Flux change distributions at 3.6\,$\upmu$m for dusty disks with Ca \textsc{ii} line emission (red), versus those stars with only infrared emission (blue). Debris disks exhibiting gaseous emission have a modestly broader distribution of infrared flux changes, consistent with the simultaneous production of both dust and gas via collisions.}
\label{fig:gasvngas}
\end{figure}

\subsection{Trends across the population}
\label{sec:pop_trend}

The light curves demonstrate variability in disk emission is observed across all timescales probed, from minutes to weeks. Variability on year to decade timescales is also evident, albeit with sparse sampling, in agreement with previous studies \citep{Swan_2019, Swan_2020}. Across the available sample of debris disks with at least two IRAC epochs, 35/42 exhibit flux variability above $3\upsigma$ in at least one channel; this is $ 85^{+4}_{-7}$\,per cent, as calculated using the $\upbeta$ probability distribution corresponding to the binomial likelihood of the data with a flat prior.  Notably, all 16 disks with relatively high-cadence monitoring on day to month timescales in programme 14258 show significant variability.  Furthermore, all seven systems where disk flux changes remain below the $3\upsigma$ threshold have only two IRAC epochs each.  This strong correlation indicates it is highly probable that sparse coverage is responsible for any apparent lack of variation, and the overall results are consistent with ubiquitous variability within debris disks around white dwarfs.

Figure~\ref{fig:trends1} presents the 3.6 and 4.5\,$\upmu$m flux and colour changes for each pair of measurements plotted against each available time baseline.  Several takeaways can be seen from the figure panels: (1) larger flux changes are more likely to occur over longer baselines; (2) brightening is as common as dimming; (3) significant colour changes are rare, with no apparent correlation with time baseline; (4) disks with Ca\,{\sc ii} line emission exhibit moderately larger flux changes, but not colour changes -- the dispersion in $|\Delta F_{\rm 3.6}|$ is 9.6\,per cent for disks with gas emission, compared to 7.2\,per cent for the rest of the sample, as shown in Figure~\ref{fig:gasvngas}. These are all consistent with previous findings \citep{Swan_2020}, but are now supported by nearly three times more data, including significantly shorter timescales, than in prior studies.

To better characterise the observed infrared excesses, spectral energy distributions (SEDs) are next constructed for each target at each photometric epoch. The combined optical and infrared data are fitted using a fixed stellar model as described earlier, plus a blackbody, with parameters calculated independently at each epoch, incorporating photometry from 2MASS \citep{Skrutskie_2006} where available. {The calculation provides the best-fitting blackbody temperature $T_{\rm dust}$, which sets the characteristic dust orbital radius $R_{\rm dust}$, and determines the total emitting area from the amplitude of the excess. Together, these yield the dust disk luminosity $L_{\rm dust}$.} Correlations between these fitted dust parameters, stellar parameters, and the maximum observed variability are next investigated for the sample as a whole.  For each star, the median dust parameter values are given in Table~\ref{tab:targets}, together with the optical depth $L_{\rm dust}/L_\star$.

\begin{figure*}
\includegraphics[width=\textwidth]{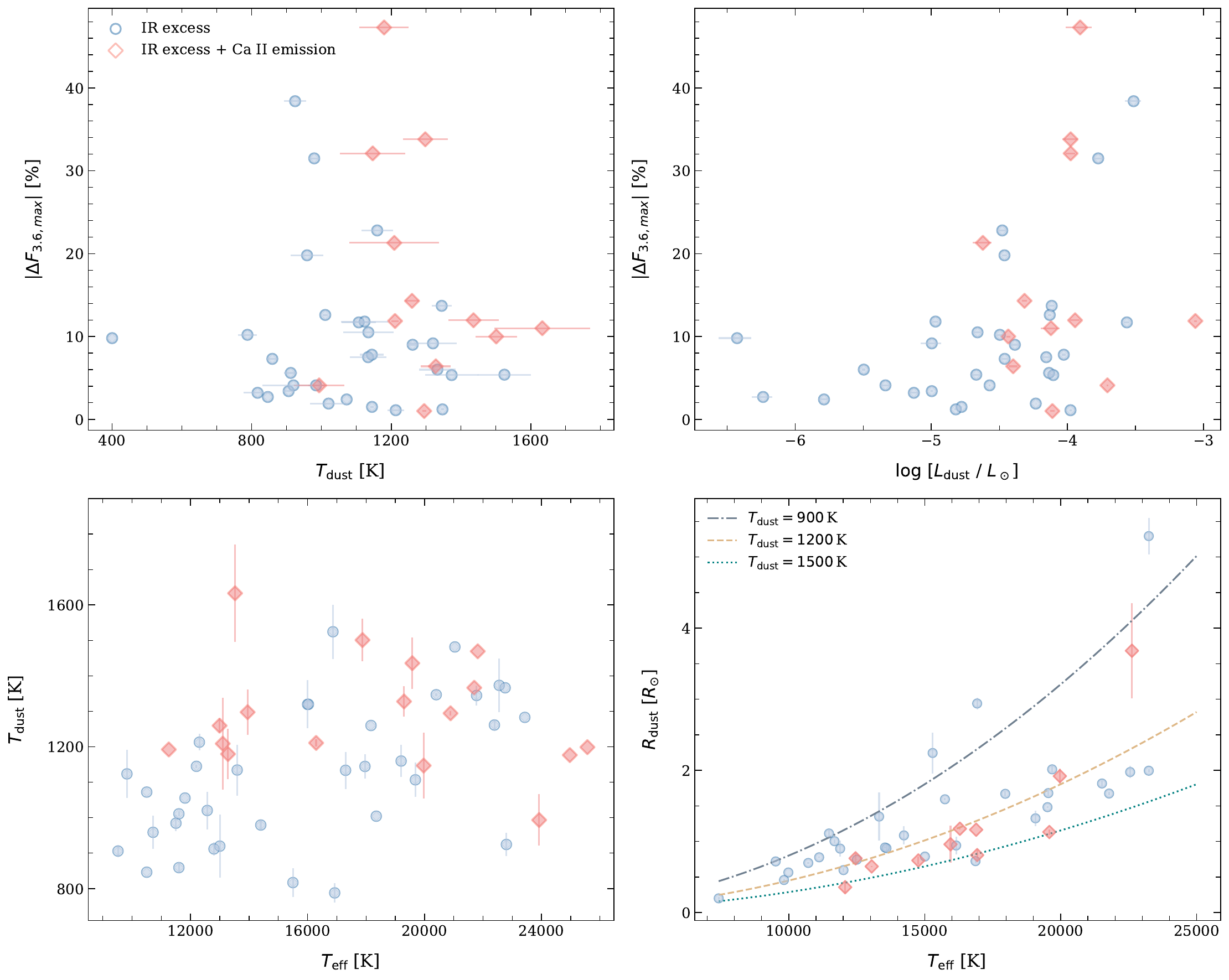}
\vskip -3pt
\caption{{A sample of plots comparing the maximum observed flux change, the fitted dust disk parameters, and stellar effective temperature.  Dusty disks with detected gas emission are plotted as red diamonds, while those with only infrared emission are indicated by blue circles. The symbols with error bars all represent the median and standard deviation of the multi-epoch disk parameters.  In the bottom-right panel, contours of constant $T_{\rm dust}$ trace the corresponding $R_{\rm dust}$ as a function of $T_{\rm eff}$, assuming a typical white dwarf radius of 0.013\,$R_{\odot}$. In these plots, 3.6\,$\upmu$m is used as a benchmark, but similar trends are found using 4.5\,$\upmu$m.  The data show that disks around warmer stars tend to have hotter dust emission, brighter disks tend to exhibit greater flux changes, and that the dust is located further out around hotter white dwarfs.}}
\label{fig:trends3}
\end{figure*}

Figure~\ref{fig:trends3} presents scatter plots for the observed maximum flux variability, the fitted dust disk parameters, and the stellar effective temperature. {As seen in the top-left panel, no correlation is observed between maximum disk variability and the median dust temperature. In the top-right panel, the positive correlation between maximum flux variability and disk luminosity might initially be attributed to the limited sampling of low-luminosity disks, particularly those with $L_{\rm dust}/L_{\odot} < 2 \times 10^{-5}$. However, Fisher’s exact test yields a p-value of 3.3\,per cent, suggesting that this trend is likely not caused by sampling bias, and that variability positively correlates with disk luminosity. In the bottom-left panel, there is tentative evidence that hotter (younger) white dwarfs tend to host disks with higher dust temperatures, with a Pearson correlation coefficient of 0.39. This is consistent with dust production at or near the Roche radius, where warmer stars would cause warmer dust emission. In the bottom-right panel, dust orbital radius is shown as a function of stellar effective temperature, with contours of constant $T_{\rm dust}$ overlaid (assuming a typical white dwarf radius of 0.013\,$R_{\odot}$). Most systems lie near the $T_{\rm dust} = 1200$\,K contour. A positive correlation is observed between $R_{\rm dust}$ and $T_{\rm eff}$, with larger orbital radii around hotter (younger) white dwarfs, consistent with rapid dust sublimation at smaller radii.} Although not shown here, no trend can be identified between disk variability and stellar parameters such as the effective temperature, mass, or spectral type. A weak positive trend is observed between disk variability and the photospheric Ca abundance when considering all systems together; however, as noted by \citet{Swan_2020}, this apparent trend arises from the distinction between systems with and without gas emission. When considered separately, neither population shows a significant correlation.

Note that it is challenging to distinguish systems with detected gas emission from disks that only manifest as infrared excesses. While gas-detected disks in this sample appear to be hotter and more luminous, small sample statistics and strong selection biases prevent any further assessment. In particular, the bulk if not all of the recently reported, gas-emission systems were identified based on sources with relatively large 3.6\,$\upmu$m excesses in {\em WISE} \citep{Melis_2020, Dennihy_2020b, GentileFusillo_2021b}, thus biasing the sample towards hotter and more luminous disks.  Future studies may be better positioned to evaluate potential differences between those with and without detectable gas, if any, beyond that highlighted in Figures~\ref{fig:gasvngas} and \ref{fig:trends3}.  No attempt is made here to search for correlations between variability and parameters of the canonical disk model, as that model suffers significant degeneracies when fitted over the wavelengths analysed here (see \citealt{Girven_2012} and references therein).

\subsection{Limits on changing dust mass}
\label{sec:dust_mass}

To better understand the observed variability, the minimum change in dust cross-sectional area $\Delta A$, and the corresponding change in dust mass can be estimated assuming optically thin, blackbody emission. The total dust cross-section is given by
\begin{equation} 
A = \int \pi r_{\rm s}^2 \, n(r_{\rm s}) \,dr_{\rm s}
\end{equation}
where $n(r_{\rm s}) = Cr_{\rm s}^{-q}$ describes the grain size distribution, spanning sizes in the range 0.1\,$\upmu$m $<r_{\rm s}<$ 1\,mm. Given an observed dust cross-section, the normalisation constant $C$ can be determined. This, in turn, allows an estimate of the total dust mass via
\begin{equation} 
M = \int \frac{4}{3} \pi r_{\rm s}^3 \uprho \,n(r_{\rm s}) \,dr_{\rm s}
\end{equation}
with $\uprho$ the dust grain density. The same approach can be applied to estimate the change in dust mass $\Delta M$, associated with a change in $A$, as inferred from the observed flux change between two epochs. These calculations yield typical changes in dust mass in the range $10^{16}$–$10^{18}$\,g across the sample, assuming $\uprho = 3$\,g\,cm$^{-3}$ and a size distribution coefficient $q = 3.5$ \citep{Dohnanyi_1969}, broadly equivalent to the mass of an asteroid with a diameter of 2--9\,km. While these values should be treated as lower limits, they nevertheless provide a useful benchmark for understanding the scale of the observed variability.

\subsection{Variability across different timescales}
\label{sec:subgroups}

As briefly mentioned above, the infrared light curves plotted capture the observational results of this survey, and demonstrate a range of brightening and dimming episodes occurring on all available timescales.  Most star-disk systems show complex behaviour, with short-term variations superimposed on broader, long-term trends.  

At the minute to hour timescales probed in this study using staring-mode, variability is evident in four of the five targets, typically at the level of a few per cent. As illustrated in Figure~\ref{fig:stare1}, the light curves display diverse behaviour, with examples of flux recovery following dips, gradual rises, and short-timescale oscillations. It is plausible that some of these patterns are part of a longer-term trend or periodic behavior. However, the short observational duration precludes the detection of any periodicities associated with orbits near the Roche limit. While the sample size is small, the fact that four out of five disks exhibit clear variability on minute to hour timescales suggests that such short-timescale fluctuations are likely widespread.  It is noteworthy that the staring-mode target 1226+110 may be as variable as the other four, but its disk is $5\times$ fainter and the resulting light curve scatter precludes a detection at a similar level (Figure~\ref{fig:stare1}).

On longer timescales of days to months, significant variability is observed in all 16 targets monitored as part of programme 14258. As can be seen in Figure~\ref{fig:maxchangevbaseline}, the largest observed fractional change in channel 1 flux is 9.7\,per cent over a 1\,d baseline (J0738+1835), increasing to 18.1\,per cent at 1\,week (1729+371) and reaching 23\,per cent over 1\,month (1226+110). A similar trend is seen for flux changes in channel 2. It has been previously shown that larger flux changes are more likely to be observed over longer baselines, based on observations spanning several months to decades \citep{Swan_2020}, and the results here demonstrate that this trend also holds for shorter baselines. As seen in the right-hand panels of Figure~\ref{fig:arch_zoom}, although variability is largely stochastic, some systems (e.g.\ 0420+520, 0842+572, 1226+110, 1536+520) show hints of possible oscillations over week-long baselines, where the flux sinks then rises, or vice versa, in both channels (which are anti-correlated for 0420+520). 

At year to decade timescales, even larger flux changes become evident, with net changes of up to 40\,per cent observed across the sample. Although sampling is sparse on longer timescales, the decay profiles of the well-sampled light curves (e.g.\ 0408$-$041, 0145+234, and 2115$-$560) can generally be approximated by a $1/t$ decay \citep{Farihi_2018, Swan_2021}, consistent with the behaviour of debris undergoing a collisional cascade (see Section~\ref{sec:cascade model}).  Whenever a decay is noted in these light curves, the fluxes always tend towards some baseline flux, never the photospheric flux \citep{Farihi_2018, Swan_2021}.  Therefore, a substantial disk of $T\sim1000$\,K debris -- detectable as infrared excesses at 3.6 and 4.5\,$\upmu$m -- seems to persist over at least decades.

The takeaway from these findings is that disk flux changes statistically correlate with observational baseline: variability at a few per cent level is common across all timescales, whereas changes exceeding 10\,per cent (statistically) require baselines of months or longer.  These results are consistent with variability driven by stochastic processes (see Section~\ref{sec:flux changes}).

\section{Data Analyses and Discussion}
\label{sec:Discussion}

In this section, the flux and colour changes are directly compared with predictions of the canonical, flat disk model.  This is followed by a brief discussion of the flux variations within the context of tidally disrupted planetesimals and collisions (i.e.\ disk evolution).  A simple model is employed to translate the observed flux increases and decreases, as a function of time, into constraints for the grain size distribution in each disk.

\subsection{Incompatibility of the canonical model}
\label{sec:canonicalmodel}

Several previous studies have commented on the limitations of the canonical disk model to account for a variety of dusty white dwarf observations, including infrared variability \citep{Farihi_2017,Bonsor_2017,Swan_2020}.  Here, for the first time, there are sufficient data for a numerical comparison with this model.  The flux from a sufficiently vertically thin, optically thick disk can be calculated as
\begin{equation}
    F = \frac{2\pi \cos{i}} {D^2} \int^{R_{\rm out}}_{R_{\rm in}} B_{\upnu}(T) \,R \, dR
\end{equation}
where $B_{\upnu}$ is the Planck function, $D$ is the distance to the system, and three disk parameters: {the inclination $i$, the inner and outer radii $R_{\rm in}$ and $R_{\rm out}$, which fix $T_{\rm in}$ and $T_{\rm out}$ using}
\begin{equation}
    T(R) = \left(\frac{2}{3\pi}\right)^{1/4} \left( \frac{R_{\star}}{R}\right)^{3/4} T_{\rm eff}
\end{equation} 
\citep{Jura_2003}. For any disk of fixed inclination, the thermal emission depends entirely on $R_{\rm in}$ and $R_{\rm out}$, and one or both of these must change for the flux to vary.

\begin{figure}
\includegraphics[width=\columnwidth]{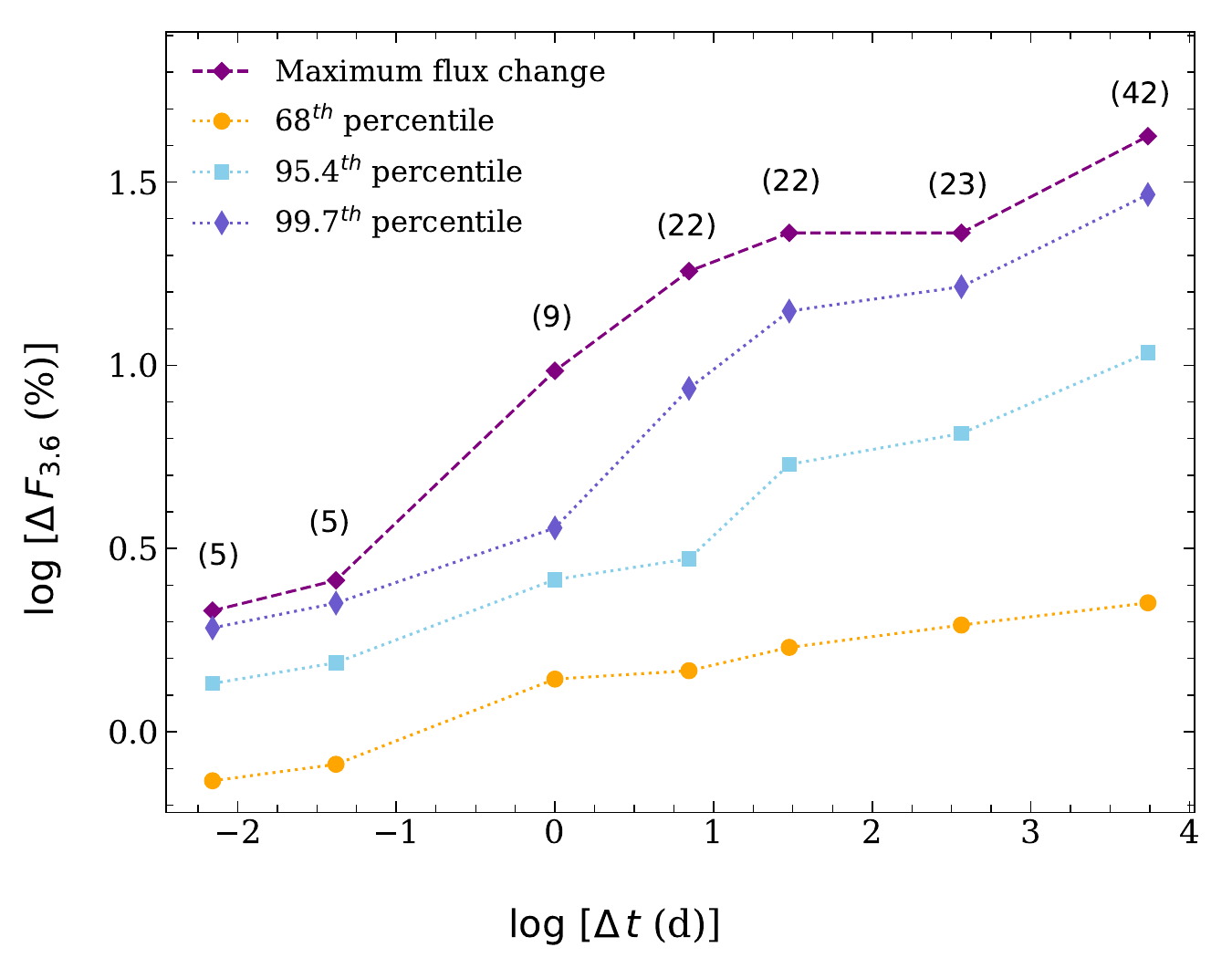}
\vskip -3pt
\caption{The plotted data points are flux changes considering all targets in the survey, for baselines of 10\,min, 1\,h, 1\,d, 7\,d, 30\,d, 1\,yr, and 15\,yr. The $68^{\rm th} (1\upsigma$), $95.4^{\rm th} (2\upsigma$), $99.7^{\rm th} (3\upsigma$) percentiles, and maximum observed flux change are shown and demonstrate that larger-amplitude changes are more likely to be observed on longer baselines. The numbers in brackets above each point indicate how many debris disks have data available at that baseline.  Analogous behaviour is seen in the 4.5\,$\upmu$m flux changes.}
\label{fig:maxchangevbaseline}
\end{figure}

\begin{figure}
\includegraphics[width=\columnwidth]{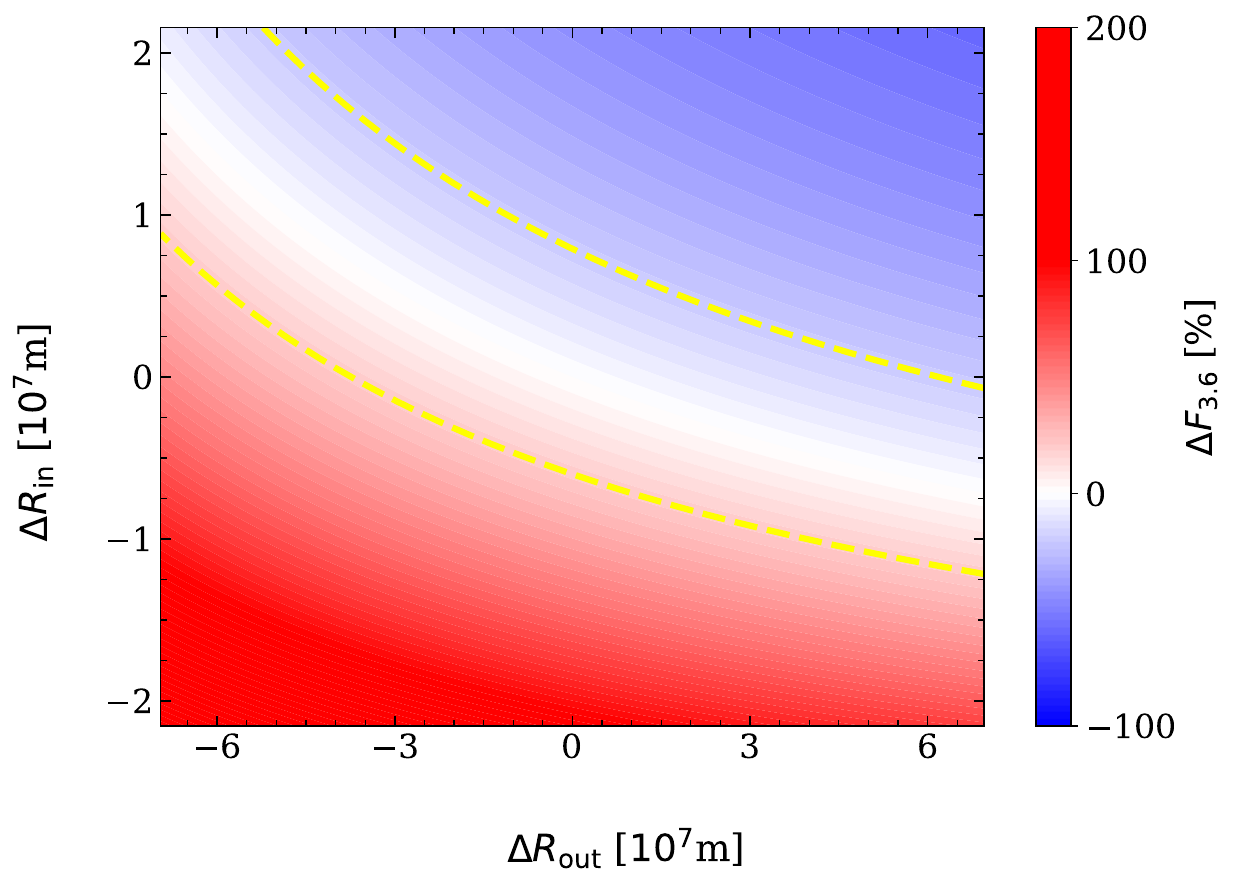}
\vskip -3pt
\caption{The predicted changes in observed 3.6\,$\upmu$m flux based on varying the inner and outer radii for the flat disk model of \citet{Jura_2003}.  The disk has a fixed inclination $i=45\degr$, and initial radii $R_{\rm in} = 0.1\,R_{\odot}$ and $R_{\rm out} = 0.3\,R_{\odot}$.  The flux changes are colour-coded within the range $-100$ to $+200$\,per cent, and between each shade are contours of constant $\Delta F_{\rm 3.6}$. The yellow dashed contours correspond to flux changes of $\pm 20$\,per cent, which are the strongest observed on timescales of a month or longer.}
\label{fig:juradisk}
\end{figure}

Consider a dust disk extending from $R_{\rm in} = 0.1\,R_{\odot}$, within which solids are expected to rapidly sublimate, to $R_{\rm out} = 0.3\,R_{\odot}$, as typically inferred for white dwarf dust disks \citep{Rocchetto_2015}, with a fixed inclination of $i=45^\degree$, around a white dwarf with $T_{\rm eff} = 12\,000$\,K. For these assumptions, and allowing both $R_{\rm in}$ and $R_{\rm out}$ to vary,  Figure~\ref{fig:juradisk} demonstrates that the 3.6\,$\upmu$m disk flux changes are typically about $3\times$ more sensitive to variations in the inner disk radius {than to variations in the outer disk radius}. To account for the observed flux changes of up to 20\,per cent within weeks to months (Figure~\ref{fig:maxchangevbaseline}), the canonical model requires a change on the order of $\Delta R_{\rm in}\sim 10^7$\,m (a 14\,per cent change). Model disks with larger $R_{\rm out}$ require even larger fractional changes in $R_{\rm in}$ to produce the same level of flux variability.

While a sudden change to the inner disk radius might be accounted for by a catastrophic event, such as an impact that adds or removes material at this particular radial location, the observed variability appears stochastic and ongoing, and not as a series of step functions, each corresponding to a singular event.  This suggests that frequent changes to $R_{\rm in}$ would be required, and both sources and sinks at this range of orbital radii, and is thus highly implausible from a physical standpoint in light of the model itself; a dynamically quiescent disk analogous to the rings of Saturn.

\begin{figure}
\includegraphics[width=\columnwidth]{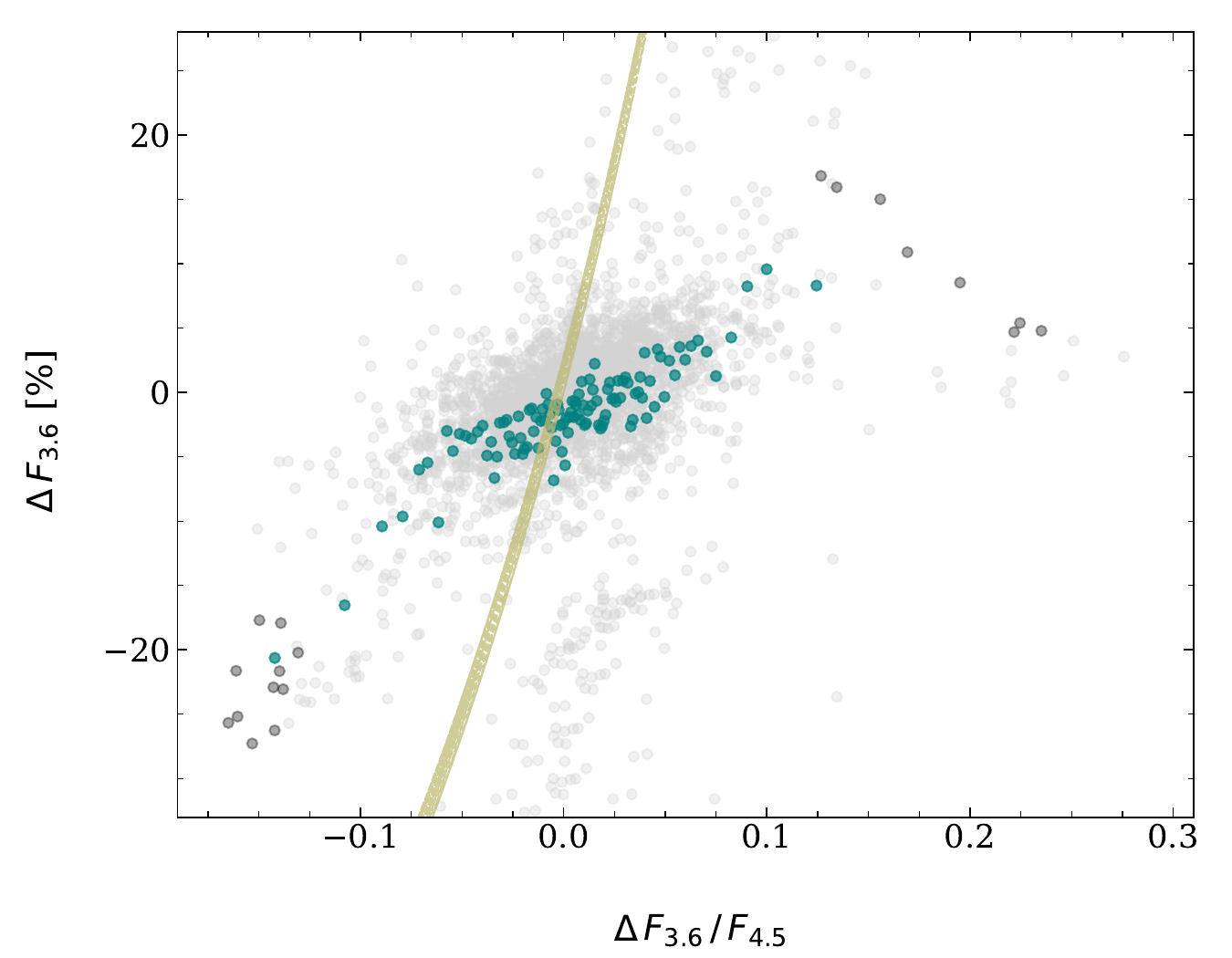}
\vskip -3pt
\caption{Epoch-wise changes in 3.6\,$\upmu$m disk flux versus colour. Grey circles represent individual measurements, with darker points indicating colour changes of $3\upsigma$ or greater. Green points show weighted averages for each evenly spaced colour bin, using 25 points per bin. Significant colour decreases are accompanied by decreases in flux, and vice versa, consistent with the systems becoming redder when dimmer, and bluer when brighter. The gold line illustrates the predictions for a flat, opaque disk, as calculated for all combinations of $\Delta R_{\rm in}$ and $\Delta R_{\rm out}$ shown in Figure~\ref{fig:juradisk}.}
\label{fig:trends2}
\end{figure}

A more decisive refutation of the canonical model can be visualised in Figure~\ref{fig:trends2}, which plots the epoch-wise flux changes at 3.6\,$\upmu$m against the corresponding colour changes for each pair of measurements for all targets. Overplotted is the expected trend for a flat, opaque disk, derived by computing the flux and colour changes for each combination of $R_{\rm in}$ and $R_{\rm out}$ spanning the plot area shown in  Figure~\ref{fig:juradisk}. {The data and the model diverge significantly in this phase space, with the model predicting colour changes that are far smaller than those observed.} Altogether, the canonical model fails to account for the observed variability, and alternative disk configurations must be considered going forward.

There is additional evidence for the limitations of the canonical model. Although not a part of this study, the growing class of transiting debris disks are similarly incompatible with a flat disk model, as the scale heights of the obscuring dust clouds must be comparable to the size of the star itself \citep{Vanderburg_2015,Vanderbosch_2021,Farihi_2022}. If the canonical model were correct, transits would not be detected, as the scale height would be far too small to induce flux drops of several per cent or more, and a flat ring would provide a constant obscuration with no variability. 

Despite these shortcomings, the flat and opaque disk model has nonetheless been foundational. Serving as the benchmark for over a decade (e.g.\ \citealt{Jura_2003, Becklin_2005, Jura_2007, Farihi_2009, Farihi_2010b, Rocchetto_2015}), it provided the first comprehensive physical framework for interpreting infrared excesses around white dwarfs, linking circumstellar dust both to tidally disrupted planetesimals and to the ongoing accretion of heavy elements in these systems. However, the accumulated observational evidence -- including the variability demonstrated here -- makes clear that this framework has now reached the limits of its explanatory power, motivating the need of a new generation of models.

\subsection{Flux variation as a result of collisional disk evolution}
\label{sec:flux changes}

The planetesimals responsible for polluting white dwarf atmospheres are thought to originate beyond a few astronomical units, owing to dynamical clearing during the first-ascent and asymptotic giant evolutionary phases of the progenitor \citep{Veras_2016}.  Thus, if the associated debris disks result from catastrophic fragmentation near the stellar Roche limit, the initial orbits must be highly eccentric \citep[e.g.][]{Debes_2012a, Veras_2014}. Tidal disruption and subsequent orbital passes can then alter the initially tight debris stream, with the subsequent evolution depending on the size of the fragmented body, its initial semimajor axis, and the orbital distance at which it experiences the strongest tidal forces \citep{Malamud_2020a, Malamud_2020b, Li_2021}.

Once disrupted, the solid debris must retain the same periastron unless influenced by non-gravitational forces, possibly resulting in repeated tidal disruptions -- and certainly collisions -- during successive close approaches \citep{Debes_2012a,Farihi_2012,Malamud2021}.  For a range of initial conditions, this complex process is predicted to generate a debris disk with a range of orbits with similar or identical periastra \citep{Malamud_2020a,Brouwers_2021}, where dust production via collisions will be enhanced at this orbit-crossing point.  Over longer timescales, additional mechanisms, such as PR drag, differential precession, or further gravitational interactions with a perturbing body, can further promote collisions \citep{Brouwers_2021, Li_2021}, thus accelerating the collisional grind-down of larger fragments into micron-sized dust.  While much of the resulting dust may remain too cool or tenuous to be detected at wavelengths beyond 10\,$\upmu$m \citep{Farihi_2014}, there must be dust sufficiently close to the star to cause $T\sim1000$\,K emission.

The link between collisions and gas production provides a cornerstone for the interpretation of the observed variability \citep[e.g.][]{Farihi_2018, Swan_2020}. The findings of this study corroborate this relatively moderate correlation, with most larger flux changes observed in disks with gaseous emission, suggesting that these are the most dynamically active disks \citep{Swan_2020}. Gas produced in collisions can rapidly (re)condense onto dust grains on timescales shorter than a single orbit \citep{Metzger_2012, Okuya_2023}, thereby serving as a potential sink for ongoing gas production.  A steady state may then exist between gas production and recondensation for most systems, as indicated by the broadly stable emission line equivalent widths. However, rare instances of significant changes in the equivalent widths of these lines have been observed over timescales of weeks or longer \citep{Wilson_2014, GentileFusillo_2021b, Rogers_2025}, possibly resulting from gas production that exceeds the capacity of the available sinks, or from changes in optical depth.

It is important to note that the variability observed at many white dwarf disks can resemble that seen in some exceptionally bright debris disks around young main-sequence stars \citep[e.g.][]{Meng_2012}. These extreme infrared excesses ($L_{\rm dust}/L_\star>1$\, per cent) are commonly accounted for by giant impacts between planetary embryos \citep[e.g.][]{Song_2005, Rhee_2008, Melis_2010b, Zuckerman_2012, Melis_2013} that can partially melt and vaporise the surfaces of the colliding bodies. These impacts place debris on a distribution of orbits, with the morphology of the resultant disk primarily shaped by the fixed collision point \citep{JacksonWyatt_2012, Jackson_2014}. As debris passes through the collision point once per orbit, it acts as an active site for dust production, and can lead to quasi-periodic variations in disk emission on orbital timescales \citep[e.g.][]{Su_2019}.

Although the physics of extreme debris disks and giant impacts may not be directly applicable to this study (e.g.\ the scale of the events), collisions between planetesimals orbiting white dwarfs remain plausible, suggesting the analogy may be relevant.  For example, while the debris generated from tidal disruption around a white dwarf can result in a range of orbits, these will intersect at periastron, establishing it as an active site for subsequent collisions. Over time, this region should remain a site of dust production and destruction as the disk evolves through further collisions. Thus, it is possible that such collisions could account for the potential short-term oscillatory-like variations observed towards some of the targets in this study (e.g.\ 0420+520, 0842+572, 1226+110, 1536+520; see Section~\ref{sec:subgroups}).

Collisions will contribute to the observed variability through cascades within a debris disk \citep{DominikDecin_2003, KenyonBromley_2004, Wyatt_2007b, Muller_2010, Gaspar_2012}. The evolution of a collisional cascade is largely dictated by the grinding down of the largest objects, which are not replenished unless a source exists. For white dwarfs, if material is supplied to the disk from an external reservoir, whether continuously or stochastically \citep{Wyatt_2014}, an equilibrium can be maintained, where the disk mass and luminosity stay roughly constant, depending on both the mass influx rate and the radius of the largest solid added to the collisional swarm \citep{KenyonBromley_2017a}. Given that planetesimal collisions offer the most plausible explanation for the observed infrared variability, the following section investigates whether the observations presented here align with the expected evolution of debris undergoing a collisional cascade, as predicted by a model.

\subsection{Application of a collisional cascade model}
\label{sec:cascade model}

In what follows, the collisional cascade model proposed by \citet{Swan_2021} -- originally developed to interpret the short-term variability in disk emission from the white dwarf 0145+234 post-outburst -- is employed to investigate the properties of white dwarf debris disks. For details of the model, the reader is referred to their work and a brief summary is provided here. The model considers the collisional evolution of optically thin debris within an annulus of width $\delta a$ at semimajor axis $a$, with vertical scale height $H$, and assumes all objects have density $\uprho = 3\,\mathrm{g\,cm}^{-3}$ (typical for solar system asteroids; \citealt{Carry_2012}). The collisional swarm is characterised by a power-law size distribution (Section~\ref{sec:dust_mass}), and particles are assumed to have sizes ranging from 0.3\,$\upmu$m to 100\,km. 

In this model, dust production is assumed to be most efficient when destructive collisions occur between objects of a similar size -- encounters with much smaller bodies are more frequent but typically non-destructive, while collisions with larger bodies are rare. To calculate these effects, the size distribution is divided into 100 logarithmically-spaced bins, and only collisions within the same bin are considered. The number of collisions between $N$ objects of size $R_{\rm s}$ within a given bin, during time interval $dt$ can be expressed as
\begin{equation}
    dN_{\rm coll} = \frac{N(N-1)}{2V} \pi R_{\rm s}^2 v_{\rm rel} dt
    \label{eq:SK}
\end{equation}
where $V= 4\pi a\delta a H$ is the volume of the annular region occupied by the debris and $v_{\rm rel}$ is the relative velocity of particles within the bin. Observed flux changes can then be used to estimate the amount of dust liberated in a single collision, and thus, the size of the colliding body.  For a given value of the power-law index $q$, this approach enables the identification of the size bin in which the relevant collisions occur. Equation~\ref{eq:SK} can then be used to determine the value of $\delta a$ that results in $dN_{\rm coll}=1$ over the observed time interval. 

In this study, the model is applied to the 16 disks observed in programme 14258 and 0145+234, i.e.\ those with good observational coverage on day to weekly timescales. Specifically, increases in the 3.6\,$\upmu$m flux between successive epochs are used to constrain the magnitude of liberated dust, while the time interval between observations informs the frequency of these collisions; this is distinct from \citet{Swan_2021}, who assumed a 1\,per cent increase on average every 2\,d for their target. The range of fractional annuli widths ($\delta a/a$) that can produce the observed flux changes is then examined for power-law slopes in the range $3 < q < 4$ (see Figure~\ref{fig:SK1}). 

For the parameters adopted here, the average vertical scale height across the sample of debris disks remains relatively large, $H/a \simeq 0.01$ (i.e.\ $H \sim 0.01\,R_{\odot}$), even for small eccentricity $e = 0.01$. This is consistent with numerical simulations that show destructive collisions near the white dwarf Roche radius cannot result in a vertically narrow (flat) disk, and only non-destructive collisions can lead to dynamical relaxation \citep{KenyonBromley_2017a}.  The modeled disk scale height can be compared to transit depths for white dwarfs with light curve dimming events that are consistent with occulations by their debris disks.  For transit depths of 3 to 30\,per cent that are seen in systems with orbital periods of hours to days \citep[e.g.][]{Guidry_2021}, these transiting clouds have been estimated to be as large or larger than the white dwarf \citep{Gansicke_2016}.  However, this is likely a significant overestimate, because individual dimming event durations are such that the obscuring dust clouds must be azimuthally extended up to $10^5$\,km in the orbital plane \citep{Farihi_2022}.  Using a rectangular approximation for these occultations, the height of the disk structures are in the range 400-4000\,km and thus in broad agreement with the model assumptions used here.

Under the assumption of a toroidal-like geometry for the white dwarf debris disks, it then follows that $H/a \sim \delta a/a$, and thus $\delta a/a\sim0.01$. The modelling results are shown in Figure~\ref{fig:SK1}, where the ranges of power-law slopes capable of producing such annuli widths are found to be approximately $3.17<q<3.34$, across the entire sample. This suggests a broadly similar size distribution among the most and least dynamically active disks, at least within the context of this model. The estimated range deviates from the canonical value of $q = 3.5$ \citep{Dohnanyi_1969} and indicates a grain size distribution that is more shallow, where the debris disks are dominated by a relatively higher proportion of larger particles, similar to what is observed in some debris disks around younger main-sequence stars \citep[e.g.][]{MacGregor_2016}.

It is important to note that the canonical value of $q = 3.5$, often cited in the context of collisional evolution, assumes constant material strength, across all object sizes. However, numerous studies have shown that material strength is size-dependent, with two distinct regimes: a strength-scaled regime in which strength increases with decreasing size, and a gravity-scaled regime in which strength increases with increasing size \citep[e.g.][]{Davis_1994, LoveAhrens_1996, HousenHolsapple_1999}. The parameter $s$ which measures the variation in strength with object size \citep{OBrien_2003}, is related to the power-law slope through $q = (7+s/3)/(2+s/3)$. For the range of power-law slopes estimated in this study, $0.4< s< 0.9$ is derived, and thus the dust-producing bodies in white dwarf debris disks fall within the gravity-scaled regime.  This is consistent with kilometre-sized or larger planetesimals, held together by self-gravity rather than material strength, and consistent with the parent body size estimates made in Section~\ref{sec:dust_mass} and immediately below.

It should be emphasised that the derived values of $q$ are highly sensitive to the assumed orbital eccentricity of  disk particles. While the adopted $e = 0.01$ is low, it is sufficient to ensure destructive collisions among the smallest particles. Given that $H \propto e$, and with the condition $H/a \sim  \delta a/a$, the inferred radial width of the annulus, and thus $q$, are directly tied to the assumed eccentricity. White dwarf debris disks are likely to have a highly eccentric origin, and if $e$ is larger than assumed here, the corresponding value of $\delta a/a$ would increase, implying that $q$ could be shallower than the range estimated here. A less steep size distribution would result in a larger value of $s$, though disks particles would remain within the gravity-scaled regime. Thus, while the adopted parameters provide a self-consistent picture of collisional activity within a vertically extended debris annulus, the inferred values of $q$, along with the associated dependent variables, should be interpreted with these caveats.

Finally, the model is employed to estimate the maximum size and mass of the debris responsible for producing the observed infrared excesses (eqs.\ 5 and 6 in \citealt{Swan_2021}). Given the derived $q$ values and the assumptions outlined previously, collisions between parent bodies with sizes typically a few times 100\,km are sufficient to generate dust that can account for the observed peak infrared excesses across the group, with derived masses scaling as $M\propto {R_{\rm s}}^{4-q}$ and falling in the range $10^{21}$–$10^{22}$\,g. The aforementioned caveats apply, as these estimates depend on model assumptions. Despite its limitations, the model suggests that these debris disks exhibit broadly similar characteristics, supporting its utility as a first-order description of collisions among planetesimals orbiting white dwarfs.

\subsection{Changes in dust temperature}
\label{sec:colour changes}

In what follows, the observed colour variations are discussed. Changes in colour typically imply changes in temperature, and thus nominally the location of circumstellar material, as might occur owing to orbital motion of inhomogeneously distributed debris (e.g.\ large clumps or clouds along an eccentric orbit). It is important to note that colour determination requires absolute fluxes, and hence calibration errors dominate the overall uncertainty. As a result, the bulk of individual colour change values are below $3\upsigma$ and thus statistically insignificant on their own.  However, as can be seen in Figure~\ref{fig:trends2}, there is clustering and correlation among the myriad measured colour changes, and this cannot by attributed to any known source of error.  The collective colour change distribution allows for a more robust interpretation.

As discussed in Section~\ref{sec:canonicalmodel}, the overall trend indicates that systems tend to appear redder when dimmer and bluer when brighter. This behaviour is also observed in the individual disk orbiting 0420+520, where the 3.6 and 4.5\,$\upmu$m fluxes are anti-correlated across all epochs, in such a way as to align with the broad trend in Figure~\ref{fig:trends2}. These changes can be readily accounted for by corresponding changes in the grain size distribution following planetesimal collisions. While collisions in a steady state produce a fixed size distribution, this can be perturbed by new dust produced in a sufficient impact.  The transient size distribution can become skewed towards smaller grains, either directly -- if the impact debris follows a steeper size distribution than the steady state -- or indirectly via recondensation of vapourised material \citep[e.g.][]{JohnsonMelosh_2014}. The emission from the resulting debris will be hotter, as smaller grains emit more efficiently at shorter wavelengths, and therefore bluer compared to the rest of the disk. As the system evolves, continued collisional processing will drive the size distribution back toward its steady-state configuration, resulting in redder emission over time.

If this interpretation is correct, then there should be plentiful small dust grains in white dwarf debris disks.  This is consistent with recent observations using the {\it James Webb Space Telescope} that have identified several new and relatively strong silicate emission features within dust orbiting white dwarfs, and which require micron-sized grains \citep{Swan_2024,Farihi_2025}.  It is also corroborated by most interpretations of the material obscuring white dwarfs during occultations consistent with circumstellar dust, using a model of an optically thick cloud for grey transits observed via multi-wavelength photometry \citep{Izquierdo_2018,Farihi_2022}, and in one case where wavelength-dependent transit depths have been observed \citep{Bhattacharjee_2025}.

\section{Conclusions}
\label{sec:Conclusions}

This study analyses all available, multi-epoch, {\it Spitzer} IRAC 3.6 and 4.5\,$\upmu$m photometry of white dwarf debris disks observed before the end of its mission. Variability above a $3\upsigma$ threshold is found for 85\,per cent (35/42) of disks, where the remaining seven systems have only two epochs each and the lack of variation can be attributed to sparse coverage.  This widespread variability is found despite sparse sampling for around two-thirds of the sample, with changes detected over timescales ranging from minutes to decades. Different systems exhibit distinct behaviour in flux changes over time, including shorter-term fluctuations superimposed on longer-term trends, and possible oscillatory-like behaviour on various timescales. 

The observed variability in disk emission is consistent with dust production and destruction driven by ongoing planetesimal collisions. A collisional cascade model is employed to explore disk properties and evolution, and to constrain grain size distributions. The infrared emission observed across the sample is used to estimate the total emitting area of the dust, and fine-tune model parameters to match the expected collision rates where necessary. Reasonable values for disk parameters, such as annuli widths and power-law size distribution coefficients, are obtained for the bulk of debris disks studied here, highlighting the explanatory power of this toy model.

Key results can be summarised as follows:
\begin{itemize}

\item Brightening and dimming episodes are equally likely across all timescales.

\item There is a clear positive correlation between flux change and time baseline between epochs.

\item The flux change distribution is moderately broader for disks with gas emission than for those with only infrared excess, indicating the former are more dynamically active and that the observed gas is produced in collisions.
        
\item There is a positive correlation between disk luminosity and maximum flux variation, indicating that brighter disks tend to be more dynamically active.
    
\item Flux and colour changes are correlated such that disks appear redder when dimmer and bluer when brighter, consistent with changes in the grain size distribution following planetesimal collisions.

\item The pioneering flat and opaque debris disk model is not compatible with any of the infrared variability reported here.

\end{itemize}

{\it Spitzer} has played a foundational role in the understanding of polluted white dwarfs and their associated debris disks, enabling the bulk of discoveries \citep[e.g.][]{Jura_2007, Farihi_2009, XuJura_2012}, the first wave of infrared characterisation \citep[e.g.][]{vonHippel_2007, Farihi_2008, Farihi_2010b, Brinkworth_2012}, emerging statistical trends \citep[e.g.][]{Girven_2012, Rocchetto_2015, Wilson_2019, Swan_2020}, and the first mid-infrared spectroscopy of these systems \citep{Reach_2005, Jura_2007b}. Its unprecedented sensitivity revealed a growing population of dusty white dwarfs, expanding the sample from a single known disk orbiting G29-38 \citep{ZuckermanBecklin_1987} to several dozen within a decade, and established the hallmark of warm dust associated with the white dwarf Roche limit and thus tidal disruption. 

As the mission drew to a close, the infrared monitoring provided by programme 14258, as presented here, has enabled constraints on both the timescale and magnitude of disk variability for a sizable sample, providing one of the few observational inputs available for theoretical models of debris disk evolution. While the canonical disk model has long served as a foundation, the observed disk behaviour is a drastic departure from its quiescent predictions. Combined with complementary datasets such as those from {\it WISE} and {\it JWST}, the observations presented here represent a legacy resource that will help to shape future observations, and theory, with a more complex and nuanced view of white dwarf debris disks.

\section*{Acknowledgements}
{The authors thank the anonymous referee for a careful reading of the manuscript.} This work is based on observations made with the {\it Spitzer Space Telescope}, which was operated by the Jet Propulsion Laboratory, California Institute of Technology under a contract with NASA.

\section*{Data Availability}
All data used in this study are available in the {\it Spitzer} Heritage Archive at \url{sha.ipac.caltech.edu/applications/Spitzer/SHA/}.



\bibliographystyle{mnras}
\bibliography{refs}


\appendix

\section{List of observations and flux densities}
\label{sec:obs_log}

Table \ref{tab:obs_log} provides all measured fluxes for all targets in all epochs.

\section{Multi-epoch light curves}
\label{sec:lightcurves}

All light curves for program 14258 are shown in Figure \ref{fig:arch_zoom}.

\section{Results of the collisional cascade model}

The modeling results of Section~\ref{sec:cascade model} are shown in Figure~\ref{fig:SK1}.

\setcounter{figure}{1}
\makeatletter 
\renewcommand{\thetable}{A\@arabic\c@figure}
\makeatother

\begin{table*}
\begin{center}	 
\caption{Observation log of {\it Spitzer} IRAC data for all known white dwarf debris disks at time of publication.}
\label{tab:obs_log}
\begin{tabular}{cccccccc} 
\hline

WD  & Date observed   & $F_{3.6}$     & $F_{4.5}$ &
WD  & Date observed   & $F_{3.6}$     & $F_{4.5}$\\
    & (MJD)           & ($\upmu$Jy)   & ($\upmu$Jy) &  
    & (MJD)           & ($\upmu$Jy)   & ($\upmu$Jy) \\

\hline

J0006+2858   & 53190.46 & 262.4 $\pm$ 6.3& 327.2 $\pm$ 7.2 &
0300$-$013   &  53999.21 & 116.97 $\pm$ 4.6 & 126.23 $\pm$ 3.98  \\

&    54096.18 &  & 291.3 $\pm$ 6.5 &
& 58089.09 & 135.19 $\pm$ 4.9&147.58 $\pm$ 4.41 \\

&    58803.92  & 249.9 $\pm$ 6.0& 279.3 $\pm$ 6.3 &
& 58817.27& 125.90 $\pm$ 4.8& 137.29 $\pm$ 4.2\\

0106$-$328   &  54822.67 & 56.5 $\pm$ 3.8 & 45.5 $\pm$ 2.6 &
& 58821.10 & 127.6 $\pm$ 4.8& 146.1 $\pm$ 4.4\\

&  56542.90 &  & 53.1 $\pm$ 2.7 &
& 58822.20  & 138.0 $\pm$ 5.0& 139.8 $\pm$ 4.3\\

&  58181.94 & 39.1 $\pm$ 3.4 & 35.4 $\pm$ 2.4& 
& 58827.45 &130.8 $\pm$ 4.9 & 154.4 $\pm$ 4.6\\

0110$-$565   &  55062.09 & 70.9 $\pm$ 2.8 & 80.2 $\pm$ 2.5 &
&  58831.39 & 134.2 $\pm$ 4.9& 138.8 $\pm$ 4.2\\

&    56646.61 & 94.3 $\pm$ 3.3 & 91.4 $\pm$ 2.8 &
&  58832.15 & 132.9 $\pm$ 4.9& 145.6 $\pm$ 4.4\\

&    58078.15 & 98.9 $\pm$ 3.4 & 102.1 $\pm$ 3.0 &
&  58836.98 & 142.7 $\pm$ 5.1& 145.8 $\pm$ 4.4\\

&    58746.92 & 96.2 $\pm$ 3.4 & 102.7 $\pm$ 3.0 &
&  58839.50 &133.4 $\pm$ 4.9 & 143.3 $\pm$ 4.3\\

&    58749.75 & 98.1 $\pm$ 3.4 & 105.0 $\pm$ 3.0 &
&  58844.17 & 135.6 $\pm$ 4.9& 147.3 $\pm$ 4.4\\

&    58755.10 & 94.6 $\pm$ 3.3 & 95.5 $\pm$ 2.8 &
&  58848.42 & 134.8 $\pm$ 4.9& 143.7 $\pm$ 4.3\\

&    58758.86 & 93.3 $\pm$ 3.3 & 95.4 $\pm$ 2.8 &
&  58851.21 & 136.2 $\pm$ 5.0& 145.8 $\pm$ 4.4\\

&    58765.37  & 98.7 $\pm$ 3.4 &100.6 $\pm$ 2.9  &
&  58854.41 & 128.0 $\pm$ 4.8& 149.7 $\pm$ 4.5\\

&    58769.54 & 97.1 $\pm$ 3.4 & 98.3 $\pm$ 2.9 &
0307+078   &   54903.11 & 9.7 $\pm$ 2.0 & 12.2 $\pm$ 1.5\\

&    58778.80 & 100.6 $\pm$ 3.6 &97.1 $\pm$ 2.9  &
& 58090.43 & 7.4 $\pm$ 2.0  & 5.0 $\pm$ 1.3 \\

&    58783.92 & 93.5 $\pm$ 3.3 & 97.2 $\pm$ 2.9 &
0347+162   & 58636.86 & 58.1 $\pm$ 1.9 & 54.5 $\pm$ 1.6 \\

&    58785.19 & 95.2 $\pm$ 3.3 & 94.3 $\pm$ 2.8 &
0408$-$041   &  53998.57 & 972 $\pm$ 21 &  1147 $\pm$ 24\\

&    58793.02  & 94.3 $\pm$ 3.3 &98.3 $\pm$ 2.9  &
&     54390.30 & 934 $\pm$ 21 &  1112 $\pm$ 24\\

&    58795.63  & 99.3 $\pm$ 3.4 & 94.8 $\pm$ 2.8 &
&     54391.77 & 922 $\pm$ 20 & 1120.14 $\pm$ 24 \\

&    58803.57 & 94.0 $\pm$ 3.3 & 100.8 $\pm$ 2.9 &
&     54396.60 & 926 $\pm$ 21 &  1106 $\pm$ 23\\

&    58808.68  & 91.0 $\pm$ 3.2 & 96.9 $\pm$ 2.9 &
&     54535.88 &  956 $\pm$ 21&  1108 $\pm$ 23\\

&    58811.01 & 89.3 $\pm$ 3.2 & 99.5 $\pm$ 2.9 &
&     56792.51 & 1037 $\pm$ 23 &  1203 $\pm$ 25\\

&    58818.42 & 91.4 $\pm$ 3.3 & 89.8 $\pm$ 2.7 &
&     57010.56 & 1005 $\pm$ 22 &  1168 $\pm$ 25\\

&    58821.43 & 95.5 $\pm$ 3.3 & 97.8 $\pm$ 2.9 & 
&     58093.32 & 880 $\pm$ 20 &  999 $\pm$ 21\\

0145+234   &  58808.32 & 1562 $\pm$ 42& 1563 $\pm$ 38 &
&     58833.15 & 777 $\pm$ 18 &  921 $\pm$ 20\\

&    58842.17 & 1323 $\pm$ 37 & 1324 $\pm$ 33 &
&     58837.33 & 788 $\pm$ 18 &  914 $\pm$ 20\\

&    58844.19 & 1337 $\pm$ 37 & 1313 $\pm$ 33 &
&     58840.30 & 795 $\pm$ 18 &  923 $\pm$ 20\\

&    58840.82 &  1351 $\pm$ 37&  1329 $\pm$ 33 &
&     58845.89  & 753 $\pm$ 17 &  911 $\pm$ 20\\

&    58839.47 & 1288 $\pm$ 36 & 1287 $\pm$ 33 &
&     58849.80 & 793 $\pm$ 18 &  926 $\pm$ 20\\

&    58836.94 & 1339 $\pm$ 37 & 1331 $\pm$ 33 &
&     58852.12 & 797 $\pm$ 18 &  919 $\pm$ 20\\

&    58833.85 & 1360 $\pm$ 38 & 1335 $\pm$ 34 &
&     58854.43  & 788 $\pm$ 18 & 942 $\pm$ 20 \\

&    58831.37 & 1403 $\pm$ 38 &  1366 $\pm$ 34&
&     58860.48 & 797 $\pm$ 18 &  921 $\pm$ 20 \\

&    58829.25 & 1419 $\pm$ 39 & 1383 $\pm$ 34 &
&    58862.42 &788 $\pm$ 18  &  919 $\pm$ 20 \\

&    58827.43 & 1416 $\pm$ 39 & 1405 $\pm$ 35 &
&     58865.73  &  778 $\pm$ 18&  917 $\pm$ 20\\

&    58822.19 & 1404 $\pm$ 38 &1432 $\pm$ 35  &
&     58869.15 & 797 $\pm$ 18 &   931 $\pm$ 20\\

&    58825.43 & 1411 $\pm$ 37 & 1403 $\pm$ 35 &
&     58873.37  &  804 $\pm$ 18 &  920 $\pm$ 20\\

&    58823.70 &  & 1447 $\pm$ 36 &
&     58659.50  &   & 964 $\pm$ 21 \\

&    58820.22 & 1462 $\pm$ 40 & 1459 $\pm$ 36 &
0420+520   &  55495.52& 320.7 $\pm$ 9.1 &  \\

&    58816.96 & 1522 $\pm$ 41 & 1501 $\pm$ 37 &
&       58106.81& 302.5 $\pm$ 8.7 & 287.0 $\pm$ 7.5 \\

0146+187   &  53960.81 & 341.1 $\pm$ 9.7 & 417 $\pm$ 10 &
&       58649.11& 350.7 $\pm$ 9.7 &  332.3 $\pm$ 8.4\\

&     58088.77 &337.0 $\pm$ 9.6  & 410 $\pm$ 10 &
&       58650.74& 352.6 $\pm$ 9.7 & 329.5 $\pm$ 8.3 \\

&     58803.82 & 341.7 $\pm$ 9.7 & 415 $\pm$ 9.9 &
&       58654.44& 353.9 $\pm$ 9.8 &  325.1 $\pm$ 8.2\\

&     58808.34 & 348.7 $\pm$ 9.8 & 409 $\pm$ 11 &
&       58659.89& 346.2 $\pm$ 9.6 &  323.0 $\pm$ 8.2\\

&     58811.32 & 335.9 $\pm$ 9.6 &  412 $\pm$ 10&
&       58661.92& 342.8 $\pm$ 9.5 & 328.8 $\pm$ 8.3 \\

&     58816.06 & 347.8 $\pm$ 9.8 &  413 $\pm$ 10&
&       58666.35 & 341.7 $\pm$ 9.5 &  328.6 $\pm$ 8.3\\

&     58817.26  & 346.8 $\pm$ 9.8 & 420 $\pm$ 9.9 &
&       58670.37& 352.9 $\pm$ 9.7 &  330.4 $\pm$ 8.3\\

&     58820.23 & 331.5 $\pm$ 9.5 & 412 $\pm$ 10 &
&       58673.90&  368.7 $\pm$ 10 &  332.2 $\pm$ 8.4\\

&     58823.70 & 343.0 $\pm$ 9.7 & 410 $\pm$ 9.8 \\

&     58829.25  & 342.1 $\pm$ 9.7 &  420 $\pm$ 10.3\\
&     58831.38 & 345.2 $\pm$ 9.7  &  423 $\pm$ 10.3\\
&     58833.86 & 351.5 $\pm$ 9.9 &  411 $\pm$ 9.9\\
&     58836.95 & 344.9 $\pm$ 9.7  &  417 $\pm$ 10.2\\
&     58840.82 & 339.5 $\pm$ 9.6 &  413 $\pm$ 10
        \\

J0234$-$0406   & 56392.03 & 74.1 $\pm$ 2.6  &  72.8 $\pm$ 2.2
\\

  & 58618.65 & 58.3 $\pm$ 2.3 & 65.2 $\pm$ 2.0 \\
  
0246+734   & 55526.43 & 22.2 $\pm$ 1.5 & 23.0 $\pm$ 1.1 \\
&      56421.30 &  & 13.98 $\pm$ 0.97 \\
&      58091.98 & 19.3 $\pm$ 1.4 &  19.4 $\pm$ 1.1
\\
\hline
\end{tabular}
\end{center}
 \end{table*}

\addtocounter{table}{-1}
\begin{table*}
\begin{center}	 
\caption{Cont.}
\begin{tabular}{cccccccc} 
\hline

WD  & Date observed   & $F_{3.6}$     & $F_{4.5}$ &
WD  & Date observed   & $F_{3.6}$     & $F_{4.5}$\\
    & (MJD)           & ($\upmu$Jy)   & ($\upmu$Jy) &  
    & (MJD)           & ($\upmu$Jy)   & ($\upmu$Jy) \\

\hline

0420$-$731   &  58079.80& 369.9 $\pm$ 9.7 &  382.7 $\pm$ 9.1 &
0842+572   &  58147.53& 1470 $\pm$ 30 & 1463 $\pm$ 30 \\

&     58679.60& 333.7 $\pm$ 9.0 & 362.9 $\pm$ 8.7 &
&   58675.29 &  &  1332 $\pm$ 27\\

&     58686.00 & 336.0 $\pm$ 9.0 &  373.8 $\pm$ 9.0&
&   58675.31 &  &  1357 $\pm$ 28\\

&     58693.53& 350.1 $\pm$ 9.3 & 362.7 $\pm$ 8.7 &
&    58672.38&  1333 $\pm$ 27&  1338 $\pm$ 27\\

&     58706.26& 351.4 $\pm$ 9.3 & 382.1 $\pm$ 9.1 &
&    58676.62& 1324 $\pm$ 27 &  1310 $\pm$ 27\\

&     58707.84& 336.8 $\pm$ 9.0 & 367.0 $\pm$ 8.9 &
&    58683.48& 1331 $\pm$ 27 & 1324 $\pm$ 27 \\

&     58716.55& 350.6 $\pm$ 9.3 & 375.8 $\pm$ 9.0 &
&    58687.18& 1361 $\pm$ 28 &  1359 $\pm$ 28\\

&     58717.63& 340.1 $\pm$ 9.1 & 366.4 $\pm$ 8.8 &
&    58689.60& 1344 $\pm$ 27 &  1364 $\pm$ 28\\

&    58735.12& 348.4 $\pm$ 9.3 & 363.5 $\pm$ 8.7 &
&    58693.60& 1307 $\pm$ 27 &  1312 $\pm$ 27\\

&     58740.66& 352.0 $\pm$ 9.3 &  370.5 $\pm$ 8.9 &
&    58699.55& 1327 $\pm$ 27 &  1320 $\pm$ 27\\

&     58746.13&  361.6 $\pm$ 9.5& 360.7 $\pm$ 8.7 &
&    58702.12& 1406 $\pm$ 29 &  1391 $\pm$ 28\\

&     58751.89& 351.4 $\pm$ 9.3 & 368.1 $\pm$ 8.8 &
&    58708.01& 1323 $\pm$ 27 &  1332 $\pm$ 27\\

&     58758.90& 358.0 $\pm$ 9.4 & 373.2 $\pm$ 8.9 &
&    58709.60& 1366 $\pm$ 28 &  1350 $\pm$ 27\\

&     58765.48& 357.0 $\pm$ 9.4 & 369.5 $\pm$ 8.9 &
&    58716.02& 1356 $\pm$ 28 & 1385 $\pm$ 28 \\

&     58775.85& 344.8 $\pm$ 9.2 & 360.6 $\pm$ 8.7 &
&    58717.30& 1354 $\pm$ 28 & 1369 $\pm$ 28 \\

&     58778.85& 338.0 $\pm$ 9.0 & 362.4 $\pm$ 8.7 &
&    58675.51&  & 1334 $\pm$ 27 \\

&     58788.80&  345.6 $\pm$ 9.2& 358.1 $\pm$ 8.6 &
0843+516   &  54228.70& 89.0 $\pm$ 2.8 & 109.5 $\pm$ 2.8 \\

&     58793.06&353.6 $\pm$ 9.4  & 351.9 $\pm$ 8.5 &
&      58151.27& 89.8 $\pm$ 2.8 &  86.9 $\pm$ 2.4\\

&     58803.58& 342.2 $\pm$ 9.1 & 360.0 $\pm$ 8.7 &
&      58680.59&  93.4 $\pm$ 2.9&  89.1 $\pm$ 2.4\\

&     58808.67& 345.1 $\pm$ 9.2 &359.4 $\pm$ 8.7  &
&      58683.49& 89.1 $\pm$ 2.8 &  86.4 $\pm$ 2.4\\

&     58816.13& 341.7 $\pm$ 9.1 & 366.7 $\pm$ 8.8 &
&      58687.19& 91.5 $\pm$ 2.8 & 84.1 $\pm$ 2.3 \\

&     58827.19& 355.9 $\pm$ 9.4 & 358.8 $\pm$ 8.7 &
&      58689.59& 86.4 $\pm$ 2.7 &  89.7 $\pm$ 2.4\\

&     58831.05& 345.0 $\pm$ 9.2 & 356.7 $\pm$ 8.6 &
&      58693.58& 92.6 $\pm$ 2.9 &  87.1 $\pm$ 2.4\\

&     58837.68& 356.6 $\pm$ 9.4 & 363.1 $\pm$ 8.7 &
&      58699.56& 92.0 $\pm$ 2.8 &  89.0 $\pm$ 2.4\\

&     58848.17& 354.3 $\pm$ 9.4 & 352.8 $\pm$ 8.5 &
&      58701.40& 88.3 $\pm$ 2.8 & 85.5 $\pm$ 2.4 \\

&     58850.49& 349.4 $\pm$ 9.3 & 365.5 $\pm$ 8.8 &
&      58706.43&  93.9 $\pm$ 2.9 & 90.5 $\pm$ 2.5 \\

&     58858.46& 340.1 $\pm$ 9.1 & 361.9 $\pm$ 8.7 &
&      58708.03& 89.4 $\pm$ 2.8 & 89.5 $\pm$ 2.4 \\

&     58865.14& 351.8 $\pm$ 9.3 &369.1 $\pm$ 8.9  &
&      58713.74& 90.5 $\pm$ 2.8 & 90.8 $\pm$ 2.5 \\

&    58873.67& 353.8 $\pm$ 9.4 & 378.1 $\pm$ 9.0&
&      58716.04& 93.3 $\pm$ 2.9 &  90.1 $\pm$ 2.4\\

0435+410   & 54908.65&  & 141.3 $\pm$ 5.3 &
&      58717.31& 90.2 $\pm$ 2.8 &  85.2 $\pm$ 2.3\\

  & 56649.01& 151.6 $\pm$ 6.9 & 140.4 $\pm$ 5.3 &
J0959$-$0200   &  55202.41& 68.9 $\pm$ 1.6 & 67.8 $\pm$ 1.5 \\

&     58106.80& 148.1 $\pm$ 6.8 & 135.4 $\pm$ 5.2 &
&     56692.48& 39.8 $\pm$ 1.0 &  42.3 $\pm$ 1.0\\

&     58649.11& 140.1 $\pm$ 6.6 & 134.0 $\pm$ 5.2 &
&     58217.71& 40.7 $\pm$ 1.1 &  42.7 $\pm$ 1.1\\

&     58650.75& 139.4 $\pm$ 6.6 & 128.0 $\pm$ 5.1 &
1015+161   &  54096.43& 129.8 $\pm$ 4.2  &  125.8 $\pm$ 3.6\\

&    58654.43& 139.1 $\pm$ 6.6 &  135.8 $\pm$ 5.2&
&     56477.09&  & 110.7 $\pm$ 3.3 \\

&     58657.84& 135.6 $\pm$ 6.6 & 130.4 $\pm$ 5.1 &
&     58182.41& 116.6 $\pm$ 4 &  117.1 $\pm$ 3.4\\

&     58659.87& 139.5 $\pm$ 6.6 & 132.3 $\pm$ 5.1 &
1018+410   & 56094.99&  49.8 $\pm$ 1.7& 52.0 $\pm$ 1.5 \\

&     58662.85 & 133.7 $\pm$ 6.5 & 130.8 $\pm$ 5.1 &
&      56101.52& 49.4 $\pm$ 1.7 &  51.6 $\pm$ 1.5 \\

&    58667.84& 139.4 $\pm$ 6.6 & 130.2 $\pm$ 5.1 &
&     58172.36& 57.5 $\pm$ 1.9 &  59.2 $\pm$ 1.7\\

&     58670.35& 140.3 $\pm$ 6.6 & 126.1 $\pm$ 5.0 &
1041+091   &  54627.50& 37.1 $\pm$ 1.1 &  26.29 $\pm$ 0.78\\

&    58673.91& 141.7 $\pm$ 6.7 &  128.8 $\pm$ 5.1&
&      58223.83&  32.6 $\pm$ 1.0& 21.11 $\pm$ 0.68 \\

0510+231   &  58659.85 & 260.2 $\pm$ 7.8 & 279.4 $\pm$ 7.3 &
1054$-$226   &  58614.50& 36.5 $\pm$ 4.4 & 18.6 $\pm$ 2.8 \\

0536$-$479   &  58093.82 & 116.0 $\pm$ 4.0 & 134.2 $\pm$ 3.8 &
1116+026   & 53890.89& 275 $\pm$ 12 & 323 $\pm$ 11 \\

0644$-$035   &  58669.89 & 184.9 $\pm$ 4.7 & 215.3 $\pm$ 5.0 &
&      58209.81& 267 $\pm$ 12 & 317 $\pm$ 10 \\

J0738+1835   &  55531.04&  74.2 $\pm$ 1.8& 70.2 $\pm$ 1.6 &
1141+057   & 58244.20& 31.62 $\pm$ 0.92 & 28.45 $\pm$ 0.76 \\

&     55531.11& 65.6 $\pm$ 1.6 & 65.1 $\pm$ 1.5 &
1145+017   & 57475.92&  & 38.0 $\pm$ 1.0 \\

&     58147.54& 97.3 $\pm$ 2.2 & 97.3 $\pm$ 2.1 &
&       57475.94&  & 35.15 $\pm$ 0.98 \\

0842+231   &  54594.95&  & 214.0 $\pm$ 5.1 &
&       57476.26&  & 43.5 $\pm$ 1.1 \\

  &  58161.78& 183.5 $\pm$ 4.9 & 181.3 $\pm$ 4.4 &
&       57847.93&  &  36.4 $\pm$ 1.0 \\

&      58693.55& 215.8 $\pm$ 5.6 & 194.2 $\pm$ 4.7 &
&        57847.95&  &  41.9 $\pm$ 1.1\\

&      58696.68 & 207.8 $\pm$ 5.4  & 197.2 $\pm$ 4.8 &
&       57848.36&  &  35.9 $\pm$ 1.0\\

&      58699.62& 205.7 $\pm$ 5.4 & 195.8 $\pm$ 4.7 &
&       58233.41&  & 35.11 $\pm$ 0.98 \\

&     58702.54& 198.0 $\pm$ 5.2 & 189.2 $\pm$ 4.6 &
&       58233.43&  &  36.7 $\pm$ 1.0\\

&      58706.40& 202.6 $\pm$ 5.3 & 191.2 $\pm$ 4.6 &
&       58233.84&  &  33.95 $\pm$ 0.96\\

&      58709.13& 204.0 $\pm$ 5.3  & 192.6 $\pm$ 4.7 &
&       58246.36& 35.1 $\pm$ 1.1 &  34.74 $\pm$ 0.97\\

&      58712.33& 209.8 $\pm$ 5.5 &  192.5 $\pm$ 4.7&
1150$-$153   & 54664.43& 419 $\pm$ 10 & 526 $\pm$ 12 \\

&     58713.80& 200.6 $\pm$ 5.3  &  190.2 $\pm$ 4.6&
&      54665.25&  429 $\pm$ 10&  540 $\pm$ 12\\

&     58714.69& 207.7 $\pm$ 5.4  & 192.6 $\pm$ 4.7 &
&      54669.75& 438 $\pm$ 10 & 547 $\pm$ 12 \\

&      58715.75& 212.2 $\pm$ 5.5  & 192.8 $\pm$ 4.7 &
&      54861.22&  445 $\pm$ 11& 545 $\pm$ 12 \\

&      58717.34& 209.5 $\pm$ 5.4 & 196.0 $\pm$ 4.7 &
&      58219.70& 416 $\pm$ 10 &  499 $\pm$ 11\\

&      58717.80& 211.8 $\pm$ 5.5 & 198.1 $\pm$ 4.8 &
J1221+1245   & 55265.03 & 76.9 $\pm$ 1.8 & 78.3 $\pm$ 1.7 \\

&     &  &  & 
& 58247.56 & 75.9 $\pm$ 1.8 & 79.0 $\pm$ 1.7 \\

\hline
\end{tabular}
\end{center}
\end{table*}

\addtocounter{table}{-1}
\begin{table*}
\begin{center}	 
\caption{Cont.}
\begin{tabular}{cccccccc} 
\hline

WD  & Date observed   & $F_{3.6}$     & $F_{4.5}$ &
WD  & Date observed   & $F_{3.6}$     & $F_{4.5}$\\
    & (MJD)           & ($\upmu$Jy)   & ($\upmu$Jy) &  
    & (MJD)           & ($\upmu$Jy)   & ($\upmu$Jy) \\
    
\hline

1226+110   & 54281.35&  194.3 $\pm$ 4.6& 213.6 $\pm$ 4.7 &
J1617+1620   & 55440.34& 75.3 $\pm$ 2.1 & 69.7 $\pm$ 1.8 \\

&      56903.41& 151.8 $\pm$ 3.7 & 166.9 $\pm$ 3.8 &
&       58086.56& 87.4 $\pm$ 2.3 &  74.9 $\pm$ 1.9\\

&      58247.64& 19.9 $\pm$ 5.7 & 5.7 $\pm$ 2.3 &
1622+587   &  58642.04&  75.1 $\pm$ 2.0 &  74.9 $\pm$ 1.8\\

&      58247.58& 146.7 $\pm$ 3.6 & 172.8 $\pm$ 3.9 &
1729+371   & 53607.42& 266.7 $\pm$ 7.3 & 325.9 $\pm$ 7.8\\

&      58749.16& 152.7 $\pm$ 3.8 & 160.6 $\pm$ 3.7 &
&        54351.24& 321.7 $\pm$ 8.4 &338.2 $\pm$ 8.1\\

&     58753.59&  152.4 $\pm$ 3.8 & 161.5 $\pm$ 3.7 &
&        54352.53& 275.0 $\pm$ 7.5 &343.1 $\pm$ 8.2\\

&      58756.17& 159.0 $\pm$ 3.9 & 167.3 $\pm$ 3.8 &
&        54356.56& 274.4 $\pm$ 7.5 &337.9 $\pm$ 8.1\\

&      58758.96& 164.8 $\pm$ 4.0 & 171.9 $\pm$ 3.9 &
&        54567.21& 261.3 $\pm$ 7.2 &323.2 $\pm$ 7.8\\

&      58764.86& 143.1 $\pm$ 3.6 & 163.7 $\pm$ 3.7 &
&       58095.90& 258.9 $\pm$ 7.2 &304.4 $\pm$ 7.3\\

&      58764.88&  148.0 $\pm$ 3.7& 157.0 $\pm$ 3.6 &
&        58661.01& 267.6 $\pm$ 7.3 &307.6 $\pm$ 7.5\\

&      58769.71& 141.0 $\pm$ 3.5 &151.4 $\pm$ 3.5  &
&        58670.99& 249.7 $\pm$ 7.0 &312.9 $\pm$ 7.6\\

&      58775.88& 135.3 $\pm$ 3.4 & 148.4 $\pm$ 3.4 &
&       58676.05& 263.5 $\pm$ 7.3 &309.8 $\pm$ 7.5\\

&      58777.34& 147.5 $\pm$ 3.7 & 150.0 $\pm$ 3.5 &
&       58684.62& 264.6 $\pm$ 7.3 &326.1 $\pm$ 7.8\\

&     58778.16& 137.1 $\pm$ 3.5 &  154.4 $\pm$ 3.5&
&       58692.04& 254.0 $\pm$ 7.1 &323.0 $\pm$ 7.8\\

&      58781.43& 147.5 $\pm$ 3.7 & 158.0 $\pm$ 3.6 &
&        58698.98& 258.8 $\pm$ 7.2 &315.1 $\pm$ 7.6\\

&      58785.70& 159.3 $\pm$ 3.9 & 152.6 $\pm$ 3.5 &
&       58705.77& 277.4 $\pm$ 7.5 &316.8 $\pm$ 7.6\\

&     58786.35&  & 160.7 $\pm$ 3.7 &
&       58714.33& 253.9 $\pm$ 7.1 &311.2 $\pm$ 7.5\\

1232+563   & 56459.56& 61.1 $\pm$ 1.5 & 73.3 $\pm$ 1.6 &
&        58732.94& 258.3 $\pm$ 7.2 &313.4 $\pm$ 7.6\\

&       58181.60& 59.7 $\pm$ 1.5 & 66.4 $\pm$ 1.5 &
&        58735.40& 261.3 $\pm$ 7.2 &312.8 $\pm$ 7.6\\

1349$-$230   & 54910.31& & 111.4 $\pm$ 2.7 &
&       58742.50& 262.2 $\pm$ 7.2 &316.2 $\pm$ 7.6\\

&       58253.79& 117.1 $\pm$ 3.1 &  92.7 $\pm$ 2.4&
&       58747.08& 262.0 $\pm$ 7.2 &318.7 $\pm$ 7.7\\

1457$-$086   & 53959.59& 40.8 $\pm$ 2.3 & 28.0 $\pm$ 1.5 &
&       58756.06& 263.4 $\pm$ 7.3 &310.2 $\pm$ 7.5\\

&       58076.73& 39.4 $\pm$ 2.3 & 26.8 $\pm$ 1.5  &
&       58766.23& 267.6 $\pm$ 7.3 &315.4 $\pm$ 7.6\\

1536+520   & 58244.13& 185.7 $\pm$ 4.1 & 196.2 $\pm$ 4.2 &
&       58775.75& 272.0 $\pm$ 7.4 &309.5 $\pm$ 7.5\\

&        58650.72& 182.2 $\pm$ 4.1 & 196.8 $\pm$ 4.2 &
&        58778.02& 264.7 $\pm$ 7.3 &334.9 $\pm$ 8.0\\

&       58655.67& 184.0 $\pm$ 4.1 & 200.4 $\pm$ 4.3 &
&        58785.72& 258.2 $\pm$ 7.2 &312.3 $\pm$ 7.6\\

&        58662.14& 183.5 $\pm$ 4.1 & 201.0 $\pm$ 4.3 &
&       58795.94& 269.7 $\pm$ 7.4 &328.9 $\pm$ 7.9\\

&        58668.26& 181.3 $\pm$ 4.0 & 198.9 $\pm$ 4.2 &
&        58803.95& 266.5 $\pm$ 7.3 &314.8 $\pm$ 7.6\\

&        58674.78& 182.4 $\pm$ 4.1 & 195.9 $\pm$ 4.2 &
&        55064.86&  &326.1 $\pm$ 7.8\\

&        58680.61& 194.6 $\pm$ 4.3 & 202.6 $\pm$ 4.3 &
&       58809.85& 265.6 $\pm$ 7.3 &319.6 $\pm$ 7.7\\

&        58687.55& 180.8 $\pm$ 4.0 & 202.4 $\pm$ 4.3 &
&       58818.20& 255.3 $\pm$ 7.1 &322.8 $\pm$ 7.8\\

&       58680.61& 186.9 $\pm$ 4.2 & 199.1 $\pm$ 4.3 &
&       58827.41& 264.4 $\pm$ 7.3 &312.7 $\pm$ 7.6\\

&        58691.20& 190.5 $\pm$ 4.2 & 207.5 $\pm$ 4.4 &
&        58836.90& 258.9 $\pm$ 7.2 &308.8 $\pm$ 7.5\\

&        58697.48& 185.9 $\pm$ 4.1 & 202.3 $\pm$ 4.3 &
&        58838.64& 267.8 $\pm$ 7.3 &313.5 $\pm$ 7.6\\

&        58710.88& 183.6 $\pm$ 4.1 & 200.1 $\pm$ 4.3 &
&        58850.41& 264.7 $\pm$ 7.3 &312.0 $\pm$ 7.6\\

&       58716.64& 189.2 $\pm$ 4.2  & 205.9 $\pm$ 4.4 & 
&        58860.53& 252.4 $\pm$ 7.0 &312.0 $\pm$ 7.5\\

&        58732.00& 180.6 $\pm$ 4.0 & 198.4 $\pm$ 4.2 & 
&        58862.20& 266.8 $\pm$ 7.3 &314.4 $\pm$ 7.6\\

&        58733.73& 181.9 $\pm$ 4.1 & 202.5 $\pm$ 4.3 &
&        58873.07& 260.9 $\pm$ 7.2 &320.8 $\pm$ 7.7\\

&        58739.59& 178.5 $\pm$ 4.0 & 191.5 $\pm$ 4.1 &
J1931+0117   & 56101.14 & 765 $\pm$ 21& 741 $\pm$ 18\\

&        58747.39& 170.8 $\pm$ 3.8 & 198.3 $\pm$ 4.2 &
& 58121.94 & 715 $\pm$ 20& 703 $\pm$ 18\\

&        58753.71& 181.6 $\pm$ 4.0 & 203.8 $\pm$ 4.3 &
2100+212   & 58735.29 &565 $\pm$ 13 &585 $\pm$ 13 \\

&        58759.09& 175.1 $\pm$ 3.9  &  194.9 $\pm$ 4.2 &
2115$-$560 &53531.57  &  & 682 $\pm$ 21\\

&        58765.24& 157.7 $\pm$ 3.6  & 190.1 $\pm$ 4.1 &
  & 53857.44 & 542 $\pm$ 22& \\

&       58770.27&  177.0 $\pm$ 4.0 & 192.9 $\pm$ 4.1 &
&  54390.69 & 563 $\pm$ 23 & 707 $\pm$ 22\\

&        58777.55& 179.6 $\pm$ 4.0 &  191.9 $\pm$ 4.1 &
&  54391.57 & 572 $\pm$ 23& 701 $\pm$ 22\\

&        58784.43& 186.8 $\pm$ 4.1 & 193.7 $\pm$ 4.1 &
&  54396.59 &564 $\pm$ 23 & 709 $\pm$ 22\\

&        58788.41& 177.9 $\pm$ 4.0 & 193.0 $\pm$ 4.1 &
&  54627.17 & 538 $\pm$ 22& 683 $\pm$ 21\\

&        58802.82& 182.2 $\pm$ 4.1 & 196.3 $\pm$ 4.2 &
&  55146.08 & &644 $\pm$ 20 \\

&        58802.82& 188.4 $\pm$ 4.2 & 197.0 $\pm$ 4.2 &
&  58120.75 & 553 $\pm$ 22& 661 $\pm$ 21\\

&        58809.13& 186.8 $\pm$ 4.1 & 202.4 $\pm$ 4.3 &
&  58714.05 & & 667 $\pm$ 21\\

&        58818.20& 184.7 $\pm$ 4.1  & 199.6 $\pm$ 4.3 &
&   58714.07& & 662 $\pm$ 21\\

1541+651   & 55567.14& 298.7 $\pm$ 8.8 & 391.0 $\pm$ 9.7 &
&  58709.11 & 547 $\pm$ 22& 663 $\pm$ 21\\

&         58175.36& 332.1 $\pm$ 9.4 & 448 $\pm$ 11 &
&  58714.28 & 535 $\pm$ 22& 639 $\pm$ 20\\

1551+175   & 55436.43& 7.77 $\pm$ 0.57 & 11.70 $\pm$ 0.50 &
&  58716.57 & 534 $\pm$ 22& 647 $\pm$ 20\\

&         58086.01& 6.86 $\pm$ 0.55 &9.45 $\pm$ 0.46&
&  58717.64 & 538 $\pm$ 22& 648 $\pm$ 20\\

1554+094   & 55066.44& 29.9 $\pm$ 0.71 & 36.70 $\pm$ 0.80 &
&  58732.66 & 544 $\pm$ 22& 657 $\pm$ 21\\

&         57172.72&  22.8 $\pm$ 0.56& 28.6 $\pm$ 0.64 &
&  58733.36 &535 $\pm$ 22 & 649 $\pm$ 20\\

&         57524.17& 25.0 $\pm$ 0.61 & 29.6 $\pm$ 0.66 &
&  58735.00 & 539 $\pm$ 22& 645 $\pm$ 20\\

&      58086.58  & 18.5 $\pm$ 0.48 & 23.6 $\pm$ 0.54 &
&  58740.55 & 556 $\pm$ 22& 660 $\pm$ 21\\

&&&&
&  58742.44 &544 $\pm$ 22 & 650 $\pm$ 20\\

&&&&
&  58746.93 & 553 $\pm$ 22& 653 $\pm$ 21\\

&&&&
&  58753.79 & 569 $\pm$ 23& 653 $\pm$ 21\\

&&&&
&  58755.08 & 545 $\pm$ 22& 646 $\pm$ 20\\

&&&&
&  58714.27 & & 657 $\pm$ 21\\

\hline
\end{tabular}
\end{center}
\end{table*}

\addtocounter{table}{-1}
\begin{table*}
\begin{center}	 
\caption{Cont.}
\begin{tabular}{cccccccc} 
\hline

WD  & Date observed   & $F_{3.6}$     & $F_{4.5}$ &
WD  & Date observed   & $F_{3.6}$     & $F_{4.5}$\\
    & (MJD)           & ($\upmu$Jy)   & ($\upmu$Jy) &  
    & (MJD)           & ($\upmu$Jy)   & ($\upmu$Jy) \\
    
\hline

2132+096   & 55545.16& 18.0 $\pm$ 2.0 & 26.4 $\pm$ 1.6 &
2326+049   & 53335.46& 6845 $\pm$ 170 & 7897 $\pm$ 180 \\

&   58156.05 &22.3 $\pm$ 2.1& 27.6 $\pm$ 1.6&
&        53726.97&  6606 $\pm$ 160 &  \\

2207+121   & 55563.82&74 $\pm$ 1.9  & 77.3 $\pm$ 1.8 &
&       54460.50& 6781 $\pm$ 160 &  7881 $\pm$ 180\\

&       58171.12& 67.0 $\pm$ 1.8 &74.2 $\pm$ 1.8  &
&        54462.17& 6892 $\pm$ 170 &  7981 $\pm$ 180\\

2212$-$135   & 58367.28 & 142.1 $\pm$ 3.2 & 148.9 $\pm$ 3.2&
&         54465.49& 7184 $\pm$ 170 &  7983 $\pm$ 180\\

2221$-$165   & 54818.88& 76.9 $\pm$ 4.0 &  90.0 $\pm$ 3.5&
&        54665.67& 6818 $\pm$ 160 & 8333 $\pm$ 190 \\

&       58161.94& 93.2 $\pm$ 4.4 &  107.1 $\pm$ 3.8 &
&        58186.77& 7662 $\pm$ 180 &  8451 $\pm$ 190\\

&       58742.45& 103.4 $\pm$ 4.6 & 111.7 $\pm$ 3.9 &
2329+407   & 58205.08& 663 $\pm$ 23 & 1049 $\pm$ 27 \\

&       58746.11& 100.2 $\pm$ 4.5 & 107.3 $\pm$ 3.8 &
&       58809.97&  & 852 $\pm$ 23 \\

&      58748.36& 101.3 $\pm$ 4.5 & 108.5 $\pm$ 3.8 &
&        58784.40& 589 $\pm$ 21 &  883 $\pm$ 24\\

&       58752.26& 101.3 $\pm$ 4.5 & 115.7 $\pm$ 4.0 &
&        58787.80& 584 $\pm$ 21 &  843 $\pm$ 23\\

&       58753.76& 98.8 $\pm$ 4.5 & 112.9 $\pm$ 3.9  &
&        58793.33& 573 $\pm$ 21 &  854 $\pm$ 23\\

&       58758.49& 100.2 $\pm$ 4.5 & 110.2 $\pm$ 3.9 &
&       58795.92& 580 $\pm$ 21 &  826 $\pm$ 23\\

&       58764.17& 94.0 $\pm$ 4.4 & 111.6 $\pm$ 3.9  &
&        58800.50& 595 $\pm$ 21 &  843 $\pm$ 23\\

&       58765.32& 102.3 $\pm$ 4.5 & 102.8 $\pm$ 3.7 &
&        58803.93& 585 $\pm$ 21 & 823 $\pm$ 23 \\

&       58767.60& 96.0 $\pm$ 4.4 & 104.7 $\pm$ 3.7 &
&        58809.09& 557 $\pm$ 21 &  821 $\pm$ 23\\

&       58769.85& 101.8 $\pm$ 4.5 & 108.7 $\pm$ 3.8 &
&       58813.00& 573 $\pm$ 21 &  838 $\pm$ 23\\

&       58775.79& 103.5 $\pm$ 4.6 & 110.0 $\pm$ 3.9 &
&        58820.21& 574 $\pm$ 21 &  841 $\pm$ 23\\

&       58776.46& 97.3 $\pm$ 4.4 & 108.0 $\pm$ 3.8 &
&        58823.69& 565 $\pm$ 21 &  823 $\pm$ 23\\

 & &  &  &
&        58827.43& 570 $\pm$ 21 & 899 $\pm$ 24 \\

 & &  &  &
&        58831.36& 569 $\pm$ 21 & 827 $\pm$ 23 \\

\hline
\end{tabular}
\end{center}
 \end{table*}

\newpage


\setcounter{figure}{0}
\makeatletter 
\renewcommand{\thefigure}{B\@arabic\c@figure}
\makeatother

\begin{figure*}
\includegraphics[width=\textwidth]{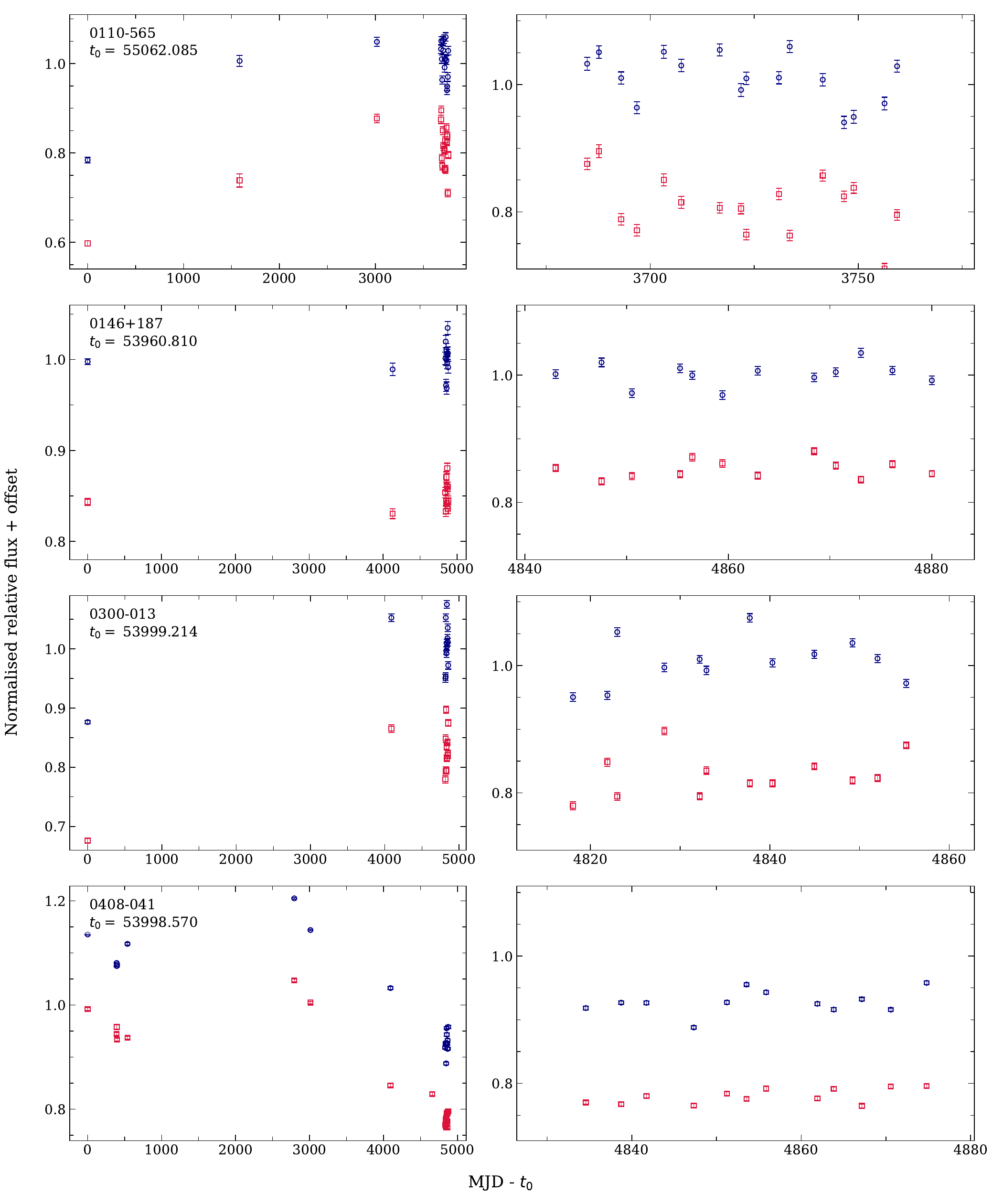}
\vskip -3pt
\caption{IRAC channels 1 and 2 light curves of all targets in programme 14258. The x-axis shows the time in days offset by a reference time $t_0$, at the start of each observation. {Blue and red circles show the normalised fluxes relative to the median flux at 3.6 and 4.5\,$\upmu$m, respectively, where the latter are offset vertically for clarity. Errors are calculated by adding the measurement errors of the target and comparison stars in quadrature.} The white dwarfs 0842+231, 0842+572, and 1226+110 show Ca \textsc{ii} line emission.}
\label{fig:arch_zoom}
\end{figure*}

\addtocounter{figure}{-1}
\begin{figure*}
\includegraphics[width=\textwidth]{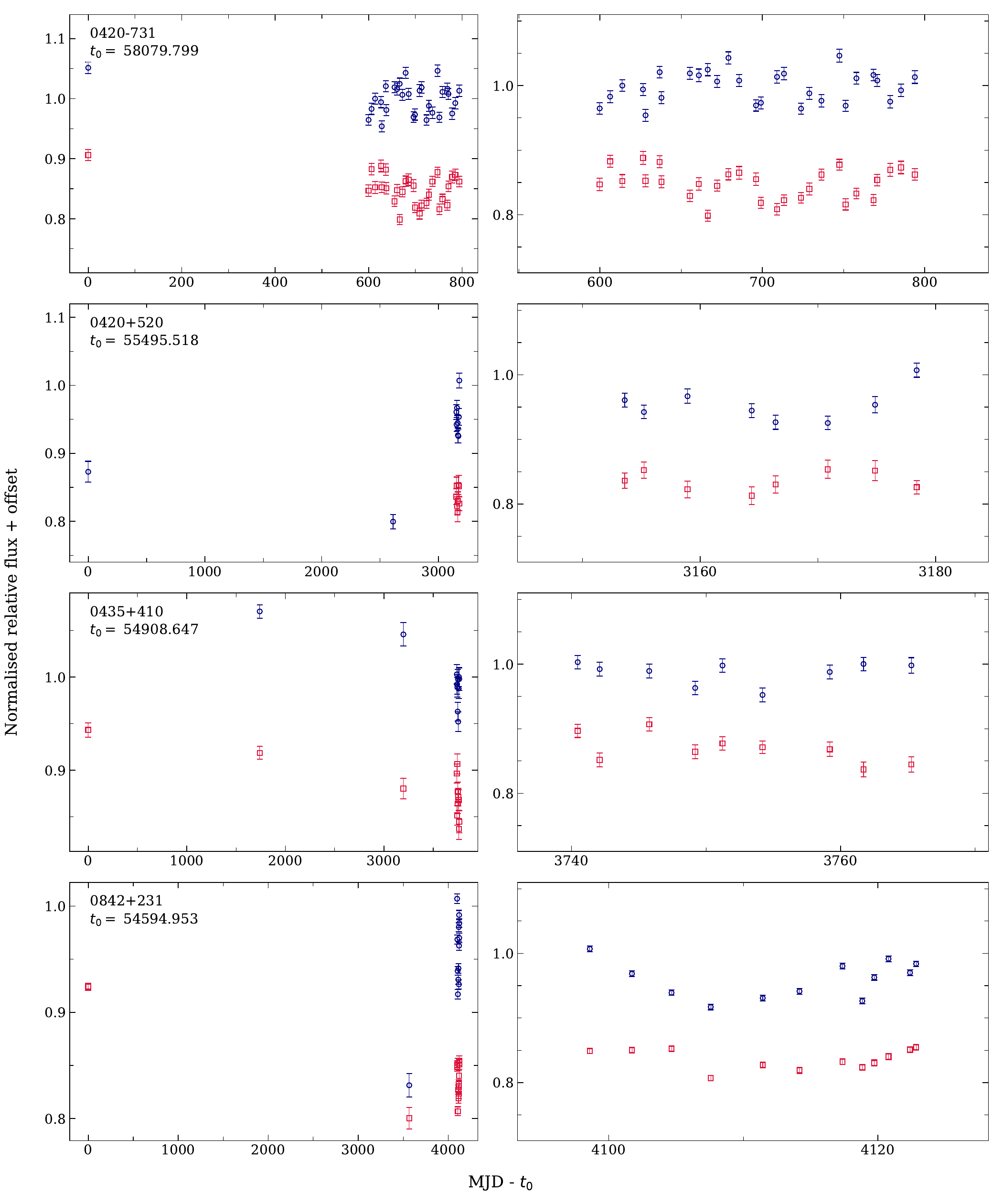}
\vskip -3pt
\caption{{\em -- continued}}
\end{figure*}

\addtocounter{figure}{-1}
\begin{figure*}
\includegraphics[width=\textwidth]{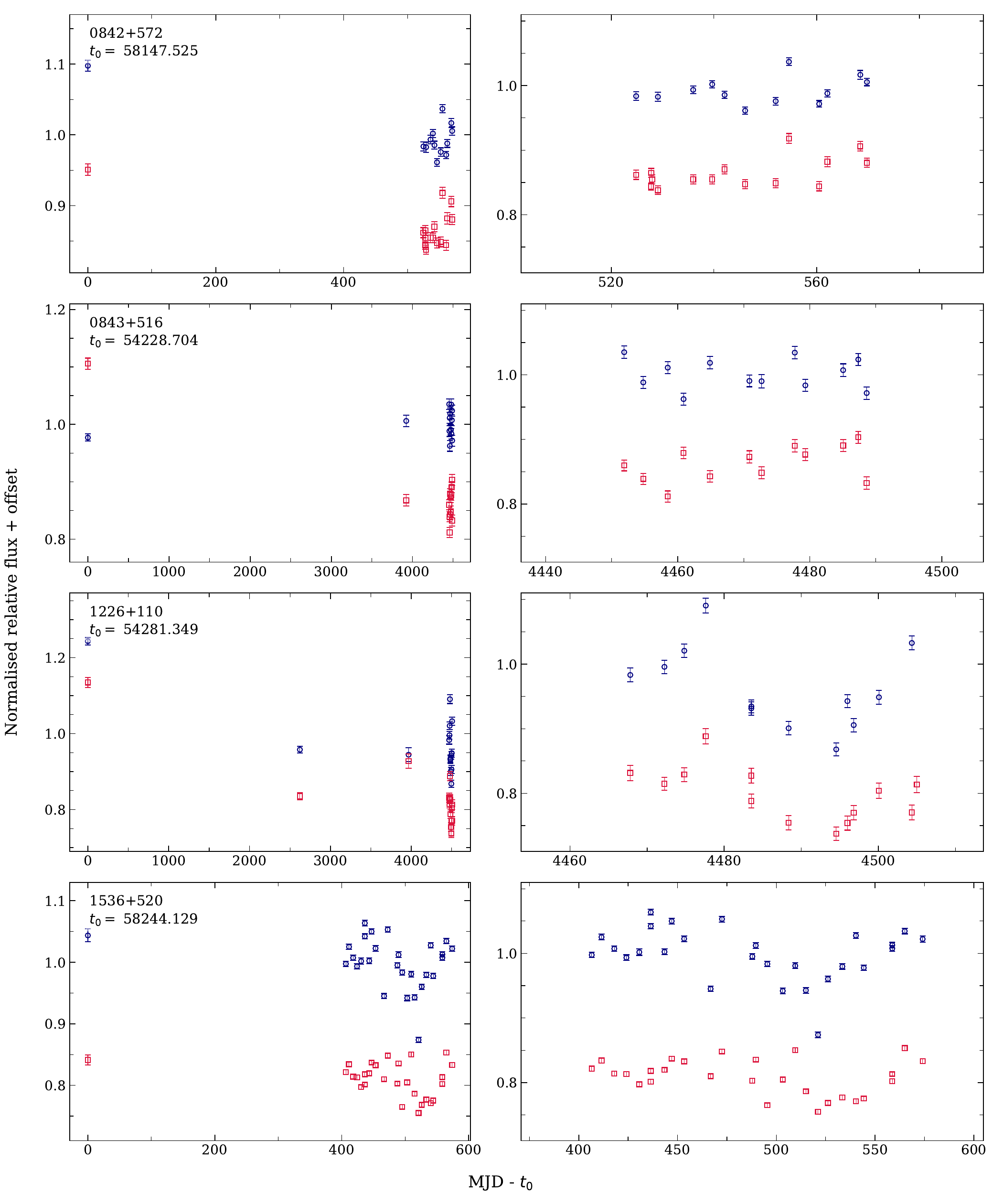}
\vskip -3pt
\caption{{\em -- continued}}
\end{figure*}

\addtocounter{figure}{-1}
\begin{figure*}
\includegraphics[width=\textwidth]{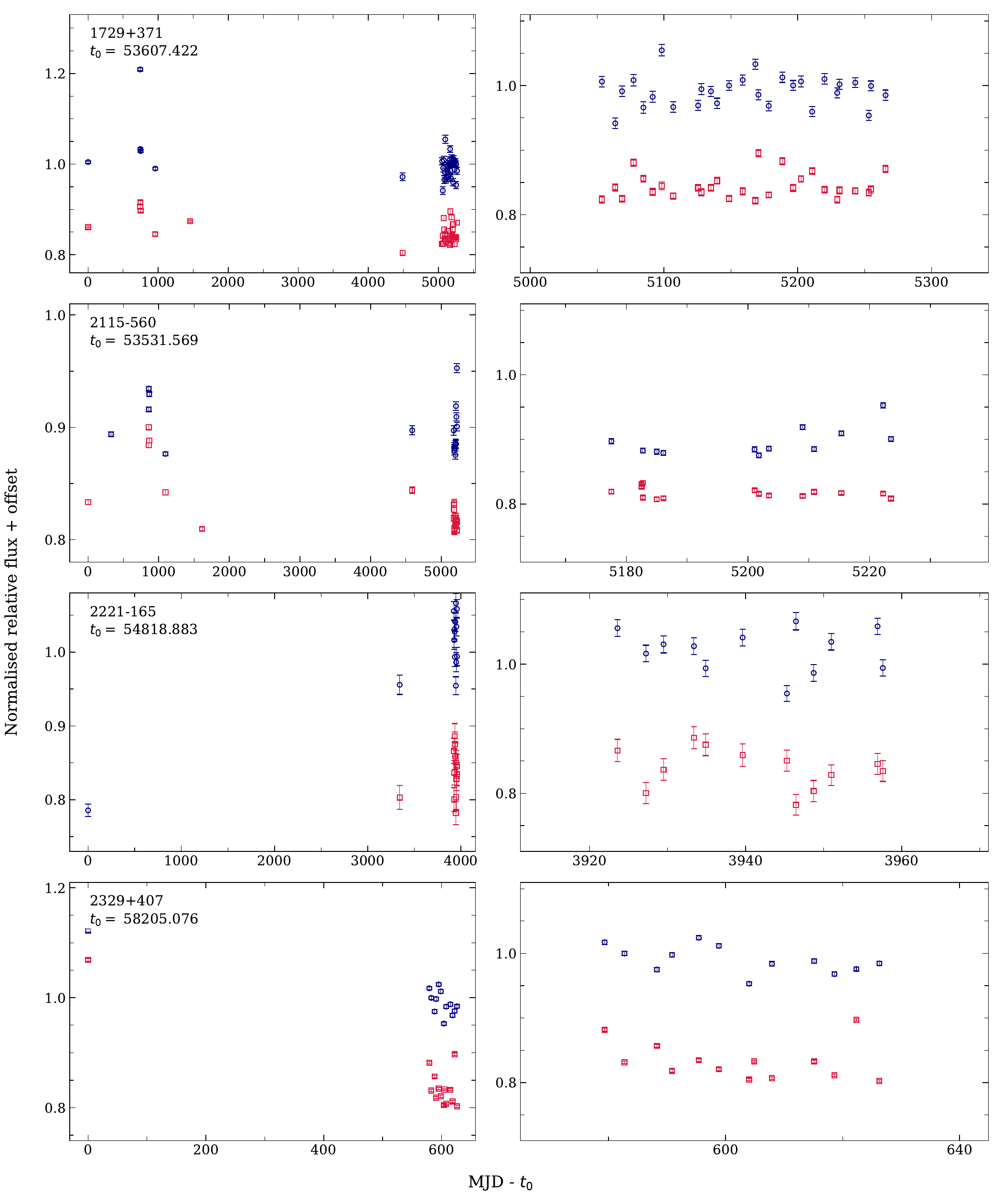}
\vskip -3pt
\caption{{\em -- continued}}
\end{figure*}




\setcounter{figure}{0}
\makeatletter 
\renewcommand{\thefigure}{C\@arabic\c@figure}
\makeatother

\begin{figure*}
\includegraphics[width=\textwidth]{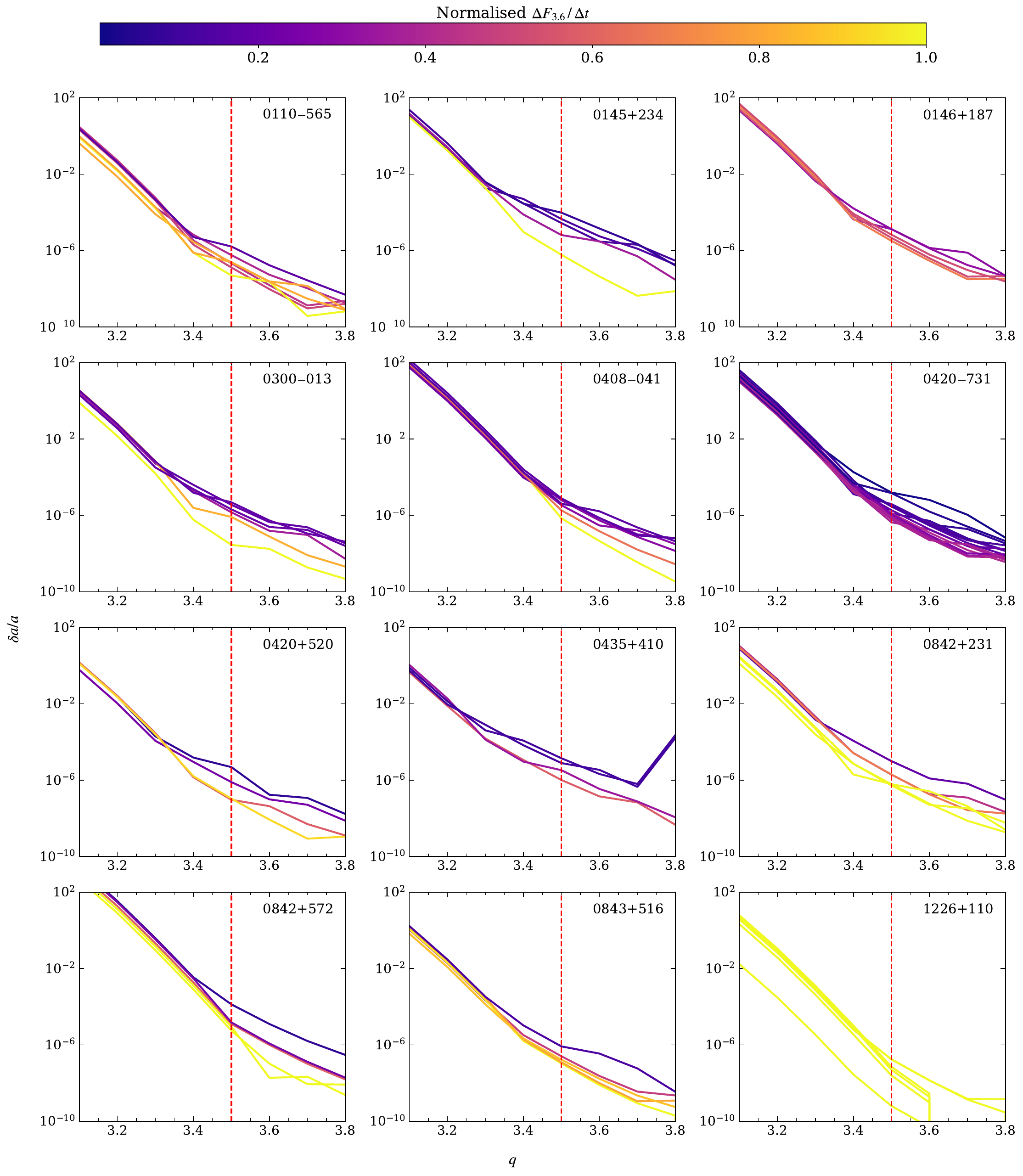}
\vskip -3pt
\caption{The range of fractional annuli width, $\delta a/a$, that can produce the observed flux changes for each target with good coverage in {\it Spitzer}, against power-law slopes in the range $3.0 < q < 3.8$. The red dashed line shows the canonical value of $q=3.5$. Lines are coloured based on the flux change in per cent divided by the time elapsed between observations in days, with lighter colours indicating larger values of this variable.}

\label{fig:SK1}
\end{figure*}

\addtocounter{figure}{-1}
\begin{figure*}
\includegraphics[width=\textwidth]{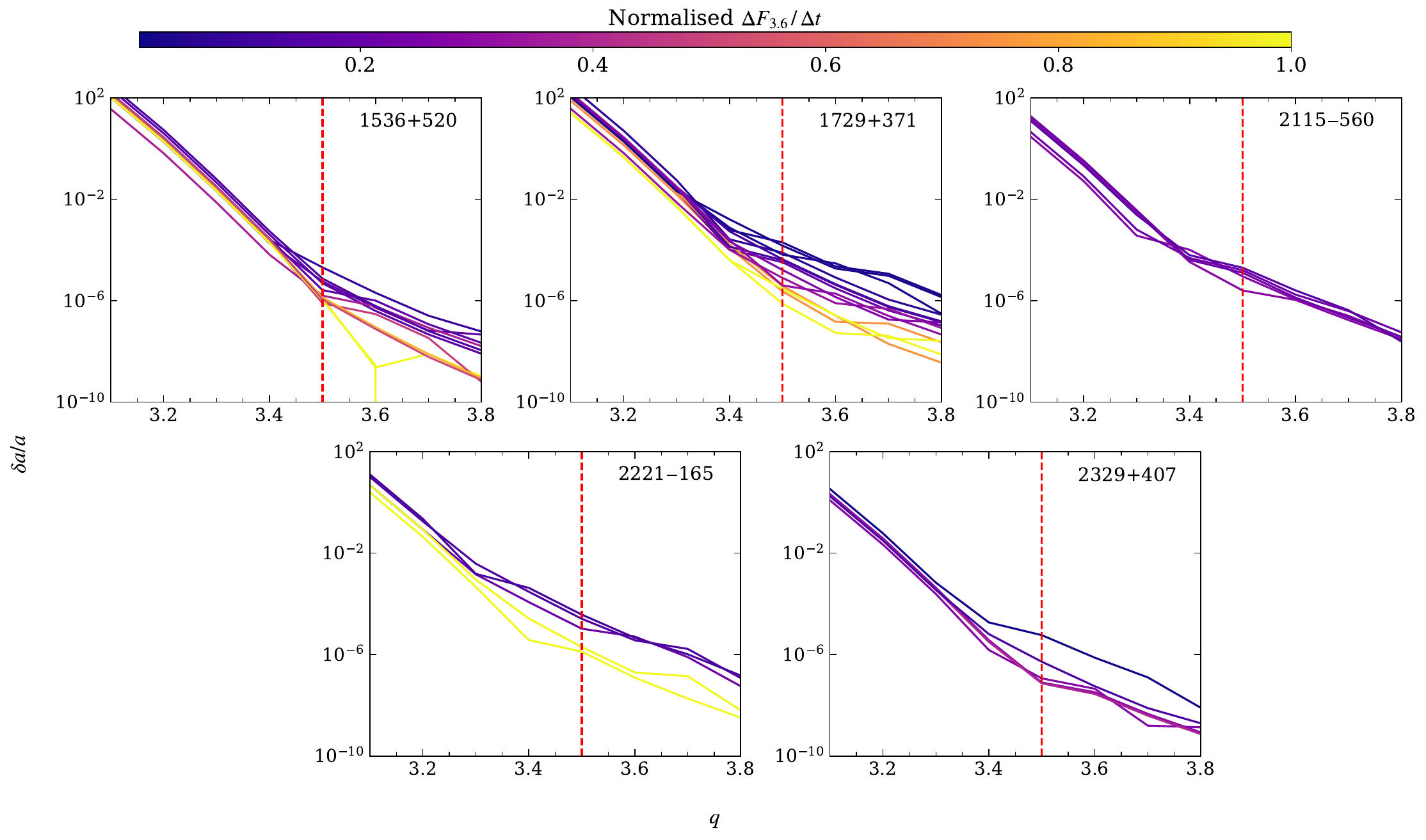}
\vskip -3pt
\caption{{\em -- continued}}
\end{figure*}

\bsp	
\label{lastpage}
\end{document}